\documentclass[10pt, final, onecolumn, twoside, romanappendices]{IEEEtran}
\usepackage{amsmath,amssymb}
\usepackage[level=3]{LatexInclusion/wgroup_message}  % set level to 0, to hide all notes
\DeclareFontFamily{OT1}{cmbr}{\hyphenchar\font45 }
\DeclareFontShape{OT1}{cmbr}{m}{n}{%
  <-9>cmbr8
  <9-10>cmbr9
  <10-17>cmbr10
  <17->cmbr17
}{}
\DeclareFontShape{OT1}{cmbr}{m}{sl}{%
  <-9>cmbrsl8
  <9-10>cmbrsl9
  <10-17>cmbrsl10
  <17->cmbrsl17
}{}
\DeclareFontShape{OT1}{cmbr}{m}{it}{%
  <->ssub*cmbr/m/sl
}{}
\DeclareFontShape{OT1}{cmbr}{b}{n}{%
  <->ssub*cmbr/bx/n
}{}
\DeclareFontShape{OT1}{cmbr}{bx}{n}{%
  <->cmbrbx10
}{}
\DeclareMathAlphabet{\mathsfbr}{OT1}{cmbr}{m}{n}%for math sans serif bright (cmbr)
\SetMathAlphabet{\mathsfbr}{bold}{OT1}{cmbr}{b}{n}%for math sans serif bright (cmbr)
\DeclareRobustCommand{\msf}[1]{%
  \ifcat\noexpand#1\relax\msfgreek{#1}\else\mathsfbr{#1}\fi%for math sans serif bright (cmbr)
}
\makeatletter
\newcommand{\msfgreek}[1]{\csname s\expandafter\@gobble\string#1\endcsname}
\makeatother

\DeclareSymbolFont{sfletters}{U}{eur}{m}{n}\SetSymbolFont{sfletters}{bold}{U}{eur}{b}{n}%for Euler Greek
\DeclareMathSymbol{\sGamma}{\mathord}{sfletters}{0}
\DeclareMathSymbol{\sDelta}{\mathord}{sfletters}{1}
\DeclareMathSymbol{\sTheta}{\mathord}{sfletters}{2}
\DeclareMathSymbol{\sLambda}{\mathord}{sfletters}{3}
\DeclareMathSymbol{\sXi}{\mathord}{sfletters}{4}
\DeclareMathSymbol{\sPi}{\mathord}{sfletters}{5}
\DeclareMathSymbol{\sSigma}{\mathord}{sfletters}{6}
\DeclareMathSymbol{\sUpsilon}{\mathord}{sfletters}{7}
\DeclareMathSymbol{\sPhi}{\mathord}{sfletters}{8}
\DeclareMathSymbol{\sPsi}{\mathord}{sfletters}{9}
\DeclareMathSymbol{\sOmega}{\mathord}{sfletters}{10}
\DeclareMathSymbol{\salpha}{\mathord}{sfletters}{"0B}
\DeclareMathSymbol{\sbeta}{\mathord}{sfletters}{"0C}
\DeclareMathSymbol{\sgamma}{\mathord}{sfletters}{"0D}
\DeclareMathSymbol{\sdelta}{\mathord}{sfletters}{"0E}
\DeclareMathSymbol{\sepsilon}{\mathord}{sfletters}{"0F}
\DeclareMathSymbol{\szeta}{\mathord}{sfletters}{"10}
\DeclareMathSymbol{\seta}{\mathord}{sfletters}{"11}
\DeclareMathSymbol{\stheta}{\mathord}{sfletters}{"12}
\DeclareMathSymbol{\siota}{\mathord}{sfletters}{"13}
\DeclareMathSymbol{\skappa}{\mathord}{sfletters}{"14}
\DeclareMathSymbol{\slambda}{\mathord}{sfletters}{"15}
\DeclareMathSymbol{\smu}{\mathord}{sfletters}{"16}
\DeclareMathSymbol{\snu}{\mathord}{sfletters}{"17}
\DeclareMathSymbol{\sxi}{\mathord}{sfletters}{"18}
\DeclareMathSymbol{\spi}{\mathord}{sfletters}{"19}
\DeclareMathSymbol{\srho}{\mathord}{sfletters}{"1A}
\DeclareMathSymbol{\ssigma}{\mathord}{sfletters}{"1B}
\DeclareMathSymbol{\stau}{\mathord}{sfletters}{"1C}
\DeclareMathSymbol{\supsilon}{\mathord}{sfletters}{"1D}
\DeclareMathSymbol{\sphi}{\mathord}{sfletters}{"1E}
\DeclareMathSymbol{\schi}{\mathord}{sfletters}{"1F}
\DeclareMathSymbol{\spsi}{\mathord}{sfletters}{"20}
\DeclareMathSymbol{\somega}{\mathord}{sfletters}{"21}
\DeclareMathSymbol{\svarepsilon}{\mathord}{sfletters}{"22}
\DeclareMathSymbol{\svartheta}{\mathord}{sfletters}{"23}
\DeclareMathSymbol{\svarpi}{\mathord}{sfletters}{"24}
\DeclareMathSymbol{\svarrho}{\mathord}{sfletters}{"25}
\DeclareMathSymbol{\svarsigma}{\mathord}{sfletters}{"26}
\DeclareMathSymbol{\svarphi}{\mathord}{sfletters}{"27}

\DeclareRobustCommand{\mcal}[1]{%
  \ifcat\noexpand#1\relax\mathnormal{#1}\else\cal{#1}\fi
}
\newcommand{\rv}[1]{\MakeLowercase{\msf{#1}}}  
\newcommand{\RV}[1]{\boldsymbol{\MakeLowercase{\msf{#1}}}}
\newcommand{\RM}[1]{\boldsymbol{\MakeUppercase{\msf{#1}}}}

\newcommand{\V}[1]{\boldsymbol{\MakeLowercase{#1}}}
\newcommand{\M}[1]{\boldsymbol{\MakeUppercase{#1}}}
\newcommand{\Set}[1]{\MakeUppercase{\mcal{#1}}}

\usepackage{hyperref} 
\hypersetup{colorlinks,citecolor= red,filecolor= blue,linkcolor= blue,urlcolor=blue}
\usepackage{acronym}  % make an acronym
\usepackage{psfrag,cite}
\usepackage{color}  % make colors
\usepackage[usenames,dvipsnames]{xcolor}
\usepackage{epstopdf}
\usepackage{calrsfs}
\usepackage{amsmath}
\usepackage{algorithm}
\usepackage{algpseudocode}
\usepackage{mathtools}
\usepackage{setspace}
\usepackage[utf8]{inputenc}
\usepackage[english]{babel}
\usepackage{amsthm}
\usepackage{physics}
\usepackage[percent]{overpic}
\usepackage{tikz}
\usepackage{pgfplots}
\usepackage{pstricks}
\usepackage{filecontents,lipsum}
\usepackage{wgroup_switch,graphicx,amsmath,amssymb}
\usepackage{tikz}
\usepackage[free-standing-units]{siunitx}
\usepackage{circuitikz}
\usepackage{float}
\usepackage{schemabloc}
\usepackage{tikz}
\usepackage{mathtools}
\usepackage{xfrac}
\usepackage{subfigure}
\usepackage{graphicx}
\usepackage{multirow}
\usepackage{pdfsync}
\usetikzlibrary{matrix}
\usetikzlibrary{circuits}
\usetikzlibrary{arrows,positioning}
\newcommand{\bd}{\begin{description}}
\newcommand{\ed}{\end{description}}
\newcommand{\be}{\begin{enumerate}}
\newcommand{\ee}{\end{enumerate}}
\newcommand{\bi}{\begin{itemize}}
\newcommand{\ei}{\end{itemize}}
\newcommand{\bl}{\begin{list}}
\newcommand{\el}{\end{list}}
\newcommand{\bt}{\begin{tabbing}}
\newcommand{\et}{\end{tabbing}}

\definecolor{BLUE}{rgb}{0,0,1}

\usepackage{acronym}  % make an acronym
\acrodef{nda}[NDA]{non-data-aided}
\acrodef{da}[DA]{data-aided}
\acrodef{crlbs}[CRLBs]{Cramer-Rao lower bounds}
\acrodef{crlb}[CRLB]{Cramer-Rao lower bound}
\acrodef{dacrlb}[DA-CRLB]{data-aided Cramer-Rao Lower Bound}
\acrodef{mds}[MDS]{maximum Doppler spread}
\acrodef{mimo}[MIMO]{multiple-input multiple-output}
\acrodef{mles}[MLEs]{maximum likelihood estimators}
\acrodef{mle}[MLE]{maximum likelihood estimator}
\acrodef{mb}[MB]{moment-based}
\acrodef{mbe}[MBE]{moment-based estimator}
\acrodef{nrmse}[NRMSE]{root-mean-square error}
\acrodef{cc}[CC]{cyclic correlation}
\acrodef{cce}[CCE]{\ac{cc} estimator}
\acrodef{mle}[MLE]{maximum likelihood estimator}
\acrodef{siso}[SISO]{single-input single-output}
\acrodef{ml}[ML]{maximum likelihood}
\acrodef{psd}[PSD]{power spectral density}
\acrodef{af}[AF]{autocorrelation function}
\acrodef{ma}[MA]{moving-average}
\acrodef{fim}[FIM]{Fisher information matrix}
\acrodef{pdf}[PDF]{probability density function}
\acrodef{mse}[MSE]{mean square error}
\acrodef{mmse}[MMSE]{minimum mean square error}
\acrodef{cdf}[CDF]{cumulative distribution function}
\acrodef{miso}[MISO]{multiple input single output}
\acrodef{ca}[CA]{code-aided}

\newcommand{\paperTitle}{{Doppler Spread Estimation in MIMO Frequency-selective Fading Channels}}

\begin{document}
\renewcommand{\figurename}{Fig.}

%%%%% UNCOMMENT THIS LINE IF YOU HAVE 2-column version %%%%%%
%\twocolumn
%%%%% UNCOMMENT THIS LINE IF YOU HAVE 2-column version %%%%%%

%---------------------------------------------------------------------------%
%                     title, title footnote, header                         %
%---------------------------------------------------------------------------%

\title{\paperTitle}

% Uncomment this line, if it's an invited paper
% \IEEEspecialpapernotice{(Invited Paper)}

% author names, IEEE memberships, corresponding address, title footnote %

\author{
%%%%%%%% uncomment this section for a 2-column formt %%%%%%%
%%%%%%%% [begin] %%%%%%%%
	\vspace{0.2cm}
%%%%%%%% [end] %%%%%%%%	
Mostafa Mohammadkarimi,~\IEEEmembership{Student Member,~IEEE,}
        Ebrahim Karami,~\IEEEmembership{Student Member,~IEEE,}
        \\
        Octavia A. Dobre, ~\IEEEmembership{Senior Member,~IEEE,}
        and Moe Z. Win~\IEEEmembership{Fellow,~IEEE}
\thanks{M. Mohammadkarimi, E. Karami, and O. A. Dobre are with the Department
of Electrical and Computer Engineering, Memorial University, St. John's, NL, Canada ({e-mail: \{\tt {m.mohammadkarimi, ekarami, odobre\}}@mun.ca}).}
\thanks{M. Z.~Win
        is with the Laboratory for Information and Decision Systems (LIDS),
        Massachusetts Institute of Technology,
        Cambridge, MA, USA
        (e-mail: {\tt moewin@mit.edu}).}}

\maketitle

\doublespacing

%---------------------------------------------------------------------------%
%                           abstract and key words                          %
%---------------------------------------------------------------------------%
\begin{abstract}
One of the main challenges in high-speed mobile communications is the presence of large Doppler spreads. Thus, accurate estimation of \ac{mds} plays an important role in improving the performance of the communication link.
{In this paper, we derive the \ac{da} and  \ac{nda} \ac{crlbs} and \ac{mles} for the \ac{mds} in \ac{mimo} frequency-selective fading channel. Moreover, a low-complexity \ac{nda}-\ac{mbe} is proposed.
The proposed \ac{nda}-\ac{mbe} relies on the second- and fourth-order moments of the received signal, which are employed to estimate the normalized squared autocorrelation function of the fading channel. Then, the problem of \ac{mds} estimation is formulated as a non-linear regression problem, and the least-squares curve-fitting optimization technique is applied to determine the estimate of the \ac{mds}.
This is the first time in the literature when \ac{da}- and \ac{nda}-\ac{mds} estimation is investigated for \ac{mimo} frequency-selective fading channel.
{Simulation results show that there is no significant performance gap between the derived \ac{nda}-MLE and
\ac{nda}-CRLB even when the observation window is relatively small.}
Furthermore, the significant reduced-complexity in the \ac{nda}-\ac{mbe} leads to low \ac{nrmse} over a wide range of \ac{mds}s when the observation window is selected large enough.}
\end{abstract}


\acresetall		% reset the acronyms

%---------------------------------------------------------------------------%
%                                Introduction                               %
%---------------------------------------------------------------------------%
\section{Introduction}\label{sec:intro}

%\mynote{Example of \texttt{mynote}}

\IEEEPARstart
{M} {aximum Doppler spread}  measures the coherence time, related to the rate of change, of wireless communication channels. Its knowledge is important to design efficient wireless communication systems for high-speed vehicles \cite{R1,R2,wu2016survey}.
In particular, accurate estimation of the \ac{mds} is required for the design of adaptive transceivers, as well
as in cellular and smart antenna systems \cite{wu2016survey,R4,R5,R6,R7,R8,R9,R10,R11,R12}. For example, in the context of adaptive transceivers,
system parameters such as coding, modulation, and power are adapted to the changes in the channel  \cite{R4,R5,R6,R7}. {In cellular systems, handoff is dictated by the velocity of the mobile station, which is also directly obtained from the Doppler information. Knowledge of the rate of the channel change is also employed to reduce unnecessary handoff; the handoff is initiated based on the received power at the mobile station, and the optimum window size for power estimation depends on the \ac{mds}
\cite{R7,R8,R9,R10}.} In the context of smart antenna systems, the \ac{mds} is used in the design of the \ac{ml} space-time transceivers \cite{R11,R12}. In addition, knowledge of \ac{mds} is required for channel tracking and equalization, as well as for the selection of the optimal interleaving length in wireless communication systems \cite{R14}.

In general, parameter estimators can be categorized as: i) \ac{da}, where the
estimation relies on a pilot or preamble sequence \cite{coleri2002channel,bellili2015maximum,marey2012based,khansefid2015channel,LiMinWin:J08}, ii) \ac{nda}, where the estimation is performed with no {\it{a priori}} knowledge about the transmitted symbols \cite{Mostafa,wang2015blind,masmoudi2011non,Mos2,stephenne2010moment}, {and
iii) \ac{ca}, where the decoding gain is used via
iterative feedback to enhance the estimation
performance of the desired parameters \cite{bellili2017time,bellili2014closed,wu2011performance,herzet2007code,simoens2006reduced,sun2005joint}.}

With regard to the \ac{mds} estimation, the \ac{da} approach often provides accurate estimates for slowly-varying channels by employing a reduced number of pilot symbols, whereas this does not hold for fast-varying channels. In the latter case,
 the details of the channel variations cannot be captured accurately, and more pilots are required, which results in increased overhead and reduced system capacity.

 There are five major classes of \ac{mds} estimators: \ac{ml}-based, \ac{psd}-based, {level-crossing-rate (LCR)-based}, covariance-based, and cyclostationarity-based estimators. The \ac{ml}-based estimator maximizes the likelihood function, and, in general, is asymptotically unbiased, achieving the \ac{crlb} \cite{R22,R17,R18}. However, \ac{mle} for \ac{mds} suffers from significant computational complexity. {Hence, different modified low-complexity \ac{mle}s for \ac{mds} in \ac{siso} flat-fading channel were developed \cite{bellili2017low,bellili2013low}.}
 With the PSD-based estimators, some unique features from the Doppler spectrum are obtained through the sample periodogram of the received signal \cite{baddour2003nonparametric}.
 Covariance-based estimators {extract} the Doppler information which exists in the sample auto-covariance of the received signal \cite{R20,R19,R225}.
 {LCR-based} estimators rely on the number of level crossings of the received signal statistics, which is proportional to the \ac{mds}\cite{park2003level}.
The cyclostationarity-based estimators exploit the cyclostationarity of the received signal \cite{R23}. Comparing with other \ac{mds} estimators, the advantage of the cyclostationarity-based estimators is the robustness to stationary noise and interference.

While the problem of \ac{mds} estimation in \ac{siso} flat-fading channel has been extensively investigated in the literature \cite{R225,bellili2017low,bellili2013low,R17,R18,R19,R20,R22,R23,baddour2003nonparametric,park2003level}, the \ac{mds} estimation in \ac{mimo} frequency-selective or in \ac{mimo} flat-fading channel has not been considerably explored.
Furthermore, \ac{da}-\ac{mds} estimation has mainly been studied in the literature.
To the best of our knowledge, only
a few works have addressed \ac{mds} etimation in conjunction with multiple antenna systems. In \cite{R18}, the authors derived an asymptotic \ac{da}-\ac{mle} and \ac{da}-\ac{crlb} for joint \ac{mds} and noise variance estimation in \ac{mimo} flat-fading channel.
In \cite{R23}, the \ac{cc} of linearly modulated signals is exploited for the \ac{mds} estimation for single transmit antenna scenarios. While both \ac{da} and \ac{nda} estimators are studied in \cite{R23}, only frequency-flat fading and single transmit antenna are considered.

{In this paper, we investigate the problem of \ac{mds} estimation in \ac{mimo} frequency-selective fading channel for both \ac{da} and \ac{nda} scenarios. The \ac{da}-\ac{crlb}, \ac{nda}-\ac{crlb}, \ac{da}-\ac{mle}, and \ac{nda}-\ac{mle} in \ac{mimo} frequency-selective fading channel are derived. In addition,
 a low-complexity \ac{nda}-\ac{mbe} is proposed.
 The proposed \ac{mbe} relies on the second- and fourth-order moments of the received signal along with the
least-square (LS) curve-fitting optimization technique to estimate the normalized squared \ac{af} and \ac{mds} of the fading channel. Since the proposed \ac{mbe} is \ac{nda}, it removes the need of pilots and preambles used for \ac{da}-\ac{mds} estimation, and thus, it results in increased system capacity. The
\ac{nda}-\ac{mbe}  outperforms the derived \ac{da}-\ac{mle}  in the presence of
 imperfect time-frequency synchronization.
Also, the \ac{mbe} outperforms the \ac{nda}-\ac{cce} in \cite{R23} and the \ac{da} low-complexity \ac{mle} in \cite{bellili2017low,bellili2013low} in \ac{siso} systems and under flat fading channels and in the presence of perfect time-frequency synchronization.}
{
\subsection{Contributions}
This paper brings the following original contributions:
\begin{itemize}
  \item The \ac{da}- and \ac{nda}-\ac{crlb}s for \ac{mds} estimation in \ac{mimo} frequency-selective fading channel are derived;
  \item The \ac{da}- and \ac{nda}-\ac{mle}s for \ac{mds} in \ac{mimo} frequency-selective fading channel are derived;
  \item  A low-complexity \ac{nda}-\ac{mbe} is proposed. The proposed estimator exhibits the following advantages:
      \begin{itemize}
      \item  lower computational complexity compared to the \ac{mle}s;
      \item  does not require time synchronization;
      \item  is robust to the carrier frequency offset;
      \item  increases system capacity;
      \item  does not require {\it a priori} knowledge of noise power, signal power, and channel delay profile;
      \item  does not require {\it a priori} knowledge of the number of transmit antennas;
      \item  removes the need of joint parameter estimation, such as carrier frequency offset, signal power, noise power, and channel delay profile estimation;
      \end{itemize}
  \item The optimal combining method for the \ac{nda}-\ac{mbe} in case of multiple receive antennas is derived through the bootstrap technique.
\end{itemize}}
\subsection{Notations}
\textit{Notation.} Random variables are displayed in sans serif, upright fonts; their realizations in serif, italic fonts. Vectors and matrices are denoted by bold lowercase and uppercase letters, respectively. For example, a random variable and its realization are denoted by $\rv{x}$ and $x$; a random vector and its realization are denoted by $\RV{x}$ and $\V{x}$; a random matrix and its realization are denoted by $\RM{X}$ and $\M{X}$, respectively.
Throughout the paper, $(\cdot)^*$ is used for
the complex conjugate, $(\cdot)^\dag$ is used for transpose,  $| \cdot |$ represents the absolute value operator, $\left\lfloor \cdot \right\rfloor$ is the floor function, $\delta_{i,j}$ denotes the Kronecker delta function, $n!$ is the factorial of $n$, $\mathbb{E}\{\cdot\}$ is the statistical expectation, $\hat{\rv{x}}$ is an estimate of $x$, and $\det(\M{A})$ denotes the determinant of the matrix $\M{A}$.

{The rest of the paper is organized as follows: Section \ref{sec:model} describes the system model;
Section \ref{section:CRLB} obtains the \ac{da}- and \ac{nda}-\ac{crlb}s for \ac{mds} estimation in \ac{mimo} frequency-selective fading;  Section \ref{section:ML} derives  the \ac{da}- and \ac{nda}-\ac{mle}s for \ac{mds} in  \ac{mimo} frequency-selective fading channel; Section \ref{section:mb} introduces the proposed \ac{nda}-\ac{mbe} for \ac{mds}; Section \ref{section:complexity} evaluates the computational complexity of the derived estimators; Section \ref{section:simulation} presents numerical results; and Section \ref{section:conclusion} concludes the paper.}

\section{System model}\label{sec:model}
Let us consider a \ac{mimo} wireless communication system with $n_{\rm{t}}$ transmit antennas and $n_{\rm{r}}$ receive antennas, where the received signals are affected by time-varying frequency-selective Rayleigh fading and are corrupted by additive white Gaussian noise.
The discrete-time complex-valued baseband signal at the $n$th receive antenna is expressed as \cite{biglieri2007mimo}
\begin{equation}\label{eq:TT1}
{\rv{r}}_k^{(n)} = \sum\limits_{m = 1}^{{n_{\rm{t}}}} {\sum\limits_{l = 1}^L {\rv{h}_{k,l}^{\left( {mn} \right)}} } \rv{s}_{k - l}^{\left( m \right)} + \rv{w}_k^{(n)}\,\,\,\,\,\,\,\,k=1,...,N,
\end{equation}
where $N$ is the number of observation symbols, $L$ is the length of the channel impulse response, $\rv{s}_k^{(m)}$ is the symbol transmitted from the $m$th antenna at time $k$, satisfying $\mathbb{E}\big{\{}\rv{s}_{k_1}^{({m_1})}{(\rv{s}_{k_2}^{({m_2})})^*}\big{\}} = \sigma _{{{\rm{s}}_{m_1}}}^2\delta_{m_1,m_2} \delta_{k_1 , k_2}$, with $\sigma _{{{\rm{s}}_{m_1}}}^2$ being the transmit power of the $m_1$th antenna, $\rv{w}_k^{(n)}$ is the complex-valued additive white Gaussian noise at the $n$th receive antenna at time $k$, whose variance is $\sigma_{{\rm{w}}_n}^2$,
 and $\rv{h}_{k,l}^{(mn)}$ denotes the zero-mean complex-valued Gaussian fading process between the $m$th transmit and $n$th receive antennas for the $l$th {tap} of the fading channel and at time $k$. It is considered that the channels for different antennas are independent, with the cross-correlation of the $l_1$ and $l_2$ {taps} given by\footnote{Here we consider the Jakes channel; it is worth noting that different parametric channel models can be also considered.}
 \begin{equation}\label{eq:TT2}
\mathbb{E}\left\{\rv{h}_{k,l_1}^{(mn)}{\big{(}\rv{h}_{k + u,l_2}^{(mn)}\big{)}^*}\right\} = \sigma _{{{\rm{h}}_{(mn),l_1}}}^2{J_0}(2\pi {f_{\rm{D}}}{T_{\rm{s}}}u)\delta_{{l_1},{l_2}},
\end{equation}
where $J_0(\cdot)$ is the zero-order Bessel function of the first kind, $\sigma_{{{\rm{h}}_{(mn),l_1}}}^2$ is the variance of the $l_1$th {tap} between
the $m$th transmit and $n$th receive antennas, $T_{\rm{s}}$ denotes the symbol period, and ${f_{\rm{D}}} = {v \mathord{\left/
 {\vphantom {v \lambda }} \right.
 \kern-\nulldelimiterspace} \lambda } = {{{f_{\rm{c}}}v} \mathord{\left/
 {\vphantom {{{f_{\rm{c}}}v} c}} \right.
 \kern-\nulldelimiterspace} c}$ represents the \ac{mds} in Hz, with $v$ as the relative speed between the transmitter and receiver, $\lambda$ as the wavelength, $f_{\rm{c}}$ as the carrier frequency, and $c$ as the speed of light.

 {
\section{CRLB for MDS Estimation}\label{section:CRLB}
In this section, the DA- and NDA-CRLB for MDS estimation in \ac{mimo} frequency-selective fading channel are derived.
\subsection{DA-CRLB}
Let us consider $\rv{s}_k^{(m)}=s_k^{(m)}$, $m=1,2,\cdots,n_{\rm{t}}$, $k=1,2,\cdots, N-L+1$, as employed pilots for DA-MDS estimation. The received signal at $n$th receive antenna in \eqref{eq:TT1} can be written as
\begin{align}\label{eq:R2} \nonumber
{\rv{r}}_k^{(n)}&=\bar{\rv{r}}_k^{(n)}+j\breve{\rv{r}}_k^{(n)}
=\sum\limits_{m = 1}^{{n_{\rm{t}}}} \sum\limits_{l = 1}^L {\bar{\rv{h}}_{k,l}^{\left( {mn} \right)}}  \bar{{s}}_{k - l}^{\left( m \right)} -{\breve{\rv{h}}_{k,l}^{\left( {mn} \right)}}  \breve{{s}}_{k - l}^{\left( m \right)}+ \bar{{w}}_k^{(n)} \\
& + j\Bigg{(}\sum\limits_{m = 1}^{{n_{\rm{t}}}} \sum\limits_{l = 1}^L {\bar{\rv{h}}_{k,l}^{\left( {mn} \right)}}  \breve{{s}}_{k - l}^{\left( m \right)} +{\breve{\rv{h}}_{k,l}^{\left( {mn} \right)}}  \bar{{s}}_{k - l}^{\left( m \right)}+ \breve{\rv{w}}_k^{(n)}\Bigg{)},
\end{align}
where $\bar{\rv{r}}_k^{(n)}\triangleq \text{Re} \big{\{}{\rv{r}}_k^{(n)}\big{\}}$, $\breve{\rv{r}}_k^{(n)}\triangleq \text{Im} \big{\{}{\rv{r}}_k^{(n)}\big{\}}$,
$\bar{\rv{h}}_{k,l}^{(mn)}\triangleq \text{Re} \big{\{}{\rv{h}}_{k,l}^{(mn)}\big{\}}$, $\breve{\rv{h}}_{k,l}^{(m,n)}\triangleq \text{Im} \big{\{}{\rv{h}}_{k,l}^{(mn)}\big{\}}$,
$\bar{{s}}_{k-l}^{(mn)}\triangleq \text{Re} \big{\{}{{s}}_{k-l}^{(mn)}\big{\}}$, and $\breve{{s}}_{k-l}^{(mn)}\triangleq \text{Im} \big{\{}{{s}}_{k-l}^{(mn)}\big{\}}$.}

{Let us define
 \begin{align} \label{eq:R3}
\RV{r}^{(n)} \triangleq \Big{[}\bar{\rv{r}}_1^{(n)} \ \bar{\rv{r}}_2^{(n)} \ \cdots \ \bar{\rv{r}}_N^{(n)} \
\breve{\rv{r}}_1^{(n)} \ \breve{\rv{r}}_2^{(n)} \ \cdots \ \breve{\rv{r}}_N^{(n)} \Big{]}^\dag
\end{align}
and
\begin{align}\label{eq:iopiop}
\RV{r} \triangleq \Big{[} {\RV{r}^{(1)}}^\dag \ {\RV{r}^{(2)}}^\dag  \ \cdots \ {\RV{r}^{(n_{\rm{r}})}}^\dag \Big{]}^\dag.
\end{align}
The elements of the vector $\RV{r}^{(n)}$,  $n=1,2,\cdots,n_{\rm{t}}$,
are linear combinations of the correlated Gaussian random variables as in \eqref{eq:R2}. Thus, $\RV{r}$,
is a Gaussian random vector with \ac{pdf} given by
\begin{align}\label{eq:R5}
p({\RV{r}}|{\V{s}};\V{\theta}) = \frac{\exp \Big{(}-\frac{1}{2}{{{\RV{r}}^{\dag}}{\M{\Sigma}^{-1}(\V{s},\V{\theta})}{\RV{r}}} \Big{)}}{{(2\pi)^{Nn_{\rm{r}}} \det^{\frac{1}{2}}  \big(\M{\Sigma}{(}\V{s},\V{\theta})\big{)}}},
\end{align}
where $\M{\Sigma}(\V{s},\V{\theta}) \triangleq \mathbb{E}{\{}\RV{r}{\RV{r}}^\dag{\}}$,
$\V{s} \triangleq \big{[} {\V{s}^{(1)}}^\dag \ {\V{s}^{(2)}}^\dag  \ \cdots \ {\V{s}^{(n_{\rm{t}})}}^\dag \big{]}^\dag$,
$\V{s}^{(m)} \triangleq  \big{[}\bar{{s}}_{1}^{(m)} \ \bar{{s}}_{2}^{(m)} \ \cdots \ \bar{{s}}_{N-L+1}^{(m)} \
\breve{{s}}_{1}^{(m)} \ \breve{{s}}_{2}^{(m)} \ \cdots \ \breve{{s}}_{N-L+1}^{(m)} \big{]}^\dag$,
and $\V{\theta} \triangleq  [\V{\xi} \  \V{\vartheta} \ f_{\rm{D}}]^\dag$ is the parameter vector, with
}
{
\begin{subequations}\label{uuuxazp}
\begin{align}
\V{\xi} &\triangleq [\sigma_{{\rm{w}}_1}^2  \ \cdots \ \sigma_{{\rm{w}}_{n_{\rm{r}}}}^2]^\dag  \\
\V{\vartheta} & \triangleq [\V{\vartheta}_1^\dag \
\V{\vartheta}_2^\dag \ \cdots \ \V{\vartheta}_L^\dag]^\dag \\
\V{\vartheta}_l & \triangleq \Big{[}{\sigma_{{{\rm{h}}_{(11),l}}}^2} \ \cdots \ {\sigma_{{{\rm{h}}_{(1n_{\rm{r}}),l}}}^2}\ {\sigma_{{{\rm{h}}_{(21),l}}}^2} \ \cdots \\ \nonumber
& \quad   \ {\sigma_{{{\rm{h}}_{(2n_{\rm{r}}),l}}}^2} \  \cdots \  {\sigma_{{{\rm{h}}_{(n_{\rm{t}}1),l}}}^2}   \cdots \ {\sigma_{{{\rm{h}}_{(n_{\rm{t}}n_{\rm{r}}),l}}}^2}\Big{]}^\dag.
\end{align}
\end{subequations}}

{Since $\RV{r}^{(n_1)}$ and
$\RV{r}^{(n_2)}$, $n_1 \neq n_2$, are uncorrelated random vectors, i.e. $\mathbb{E}\big{\{}\RV{r}^{(n_1)}{\RV{r}^{(n_2)}}^\dag\big{\}}=\M{0}$,
the covariance matrix of $\RV{r}$, $\M{\Sigma} (\V{s},\V{\theta})$, is block diagonal as
\begin{align}\label{eq:covcov}
\M{\Sigma} (\V{s},\V{\theta})\triangleq \mathbb{E}\{\RV{r}\RV{r}^{\dag}\}=
\begin{bmatrix}
  \M{\Sigma}^{(1)} &  &  &  \\
   & \M{\Sigma}^{(2)} &  &  \\
   &  & \ddots &  \\
    &  &  & \M{\Sigma}^{(n_{\rm{r}})}
\end{bmatrix},
\end{align}
where $\M{\Sigma}^{(n)} \triangleq \mathbb{E}\big{\{}\RV{r}^{(n)}{\RV{r}^{(n)}}^\dag \big{\}}$.
By employing \eqref{eq:TT2}, \eqref{eq:R2}, and \eqref{eq:R3}, using the fact the real and imaginary part of the fading tap are independent random variables with $\mathbb{E}\big{\{}|\bar{\rv{h}}_{k,l}^{(mn)}|^2\big{\}}=\big{\{}|\breve{\rv{h}}_{k,l}^{(mn)}|^2\big{\}}$
$=\sfrac{{\sigma_{{{\rm{h}}_{(mn),l}}}^2}}{2}$, and after some algebra, the elements of the covariance matrix $\M{\Sigma}^{(n)}$, $n\in \{1,2,\cdots,n_{\rm{r}}\}$, are obtained as
\begin{subequations}
\begin{align}
\mathbb{E} \Big{\{} \bar{\rv{r}}_{k}^{(n)}& \bar{\rv{r}}_{k+u}^{(n)}\Big{\}}=\mathbb{E}  \Big{\{} \breve{\rv{r}}_{k}^{(n)}  \breve{\rv{r}}_{k+u}^{(n)}\Big{\}} \\ \nonumber
&=\frac{1}{2}\sum\limits_{m = 1}^{{n_{\rm{t}}}} \sum\limits_{l = 1}^L {\sigma_{{{\rm{h}}_{(mn),l}}}^2}\Big{(} \bar{{s}}_{k-l}^{(m)}\bar{{s}}_{k+u-l}^{(m)}+\breve{{s}}_{k-l}^{(m)}
\breve{{s}}_{k+u-l}^{(m)}\Big{)} \\ \nonumber
&\,\,\,\,\,\,\,\,\,\,\,\,\,\,\,\,\,\,\,\,\,\,\,\,\,\,\,\,\,\,\,\,\ J_0(2\pi f_{\rm{D}}T_{\rm{s}}u)+\frac{\sigma_{{\rm{w}}_n}^2}{2} \delta_{u,0}
\end{align}
\begin{align}
\mathbb{E} \Big{\{} \bar{\rv{r}}_{k}^{(n)} & \breve{\rv{r}}_{k+u}^{(n)}\Big{\}}=-\mathbb{E} \Big{\{} \breve{\rv{r}}_{k}^{(n)} \bar{\rv{r}}_{k+u}^{(n)}\Big{\}} \\  \nonumber
&\frac{1}{2}\sum\limits_{m = 1}^{{n_{\rm{t}}}} \sum\limits_{l = 1}^L {\sigma_{{{\rm{h}}_{(mn),l}}}^2}\Big{(} \bar{{s}}_{k-l}^{(m)}\breve{{s}}_{k+u-l}^{(m)}-\breve{{s}}_{k-l}^{(m)}\bar{{s}}_{k+u-l}^{(m)}\Big{)} \\ \nonumber
& \quad \quad \quad \quad  \quad J_0(2\pi f_{\rm{D}}T_{\rm{s}}u).
\end{align}
\end{subequations}}

{The Fisher information matrix of the parameter vector $\V{\theta}$, $\M{I}(\V{\theta})$, for the zero-mean Gaussian observation vector in \eqref{eq:R5} is obtained as
\begin{align}
[\M{I}(\V{\theta})]_{ij} &\triangleq -\mathbb{E}\Bigg{\{}\frac{\partial^2\ln p({\RV{r}}|\V{s};\V{\theta})}{\partial \V{\theta}_i \partial \V{\theta}_j }\Bigg{\}} \\ \nonumber
& =
\frac{1}{2} {\rm{tr}} \Bigg{[}\M{\Sigma}^{-1}(\V{s},\V{\theta})\frac{\partial \M{\Sigma}(\V{s},\V{\theta})}{\partial \V{\theta}_{i}}
\M{\Sigma}^{-1}(\V{s},\V{\theta})\frac{\partial \M{\Sigma}(\V{s},\V{\theta})}{\partial \V{\theta}_{j}}
\Bigg{]}.
\end{align}
For the \ac{mds}, $f_{\rm{D}}$, ${I}(f_{\rm{D}}) \triangleq [\M{I}(\V{\theta})]_{xx}$, $x=n_{\rm{t}}n_{\rm{r}}L+n_{\rm{r}}+1$, and one obtains
\begin{align}\label{eq:R9}
{I}(f_{\rm{D}})&= -\mathbb{E}\Bigg{\{}\frac{\partial^2\ln p({\RV{r}}|\V{s};\V{\theta})}{\partial f_{\rm{D}}^2}\Bigg{\}} \\ \nonumber
&=\frac{1}{2} {\rm{tr}} \Bigg{[}\bigg{(}\M{\Sigma}^{-1}(\V{s},\V{\theta})\frac{\partial \M{\Sigma}(\V{s},\V{\theta})}{\partial f_{\rm{D}}}\bigg{)}^2\Bigg{]},
\end{align}
where $\frac{\partial \M{\Sigma}(\V{s},\V{\theta})}{\partial f_{\rm{D}}}$ is obtained by replacing $J_0(2\pi f_{\rm{D}}T_{\rm{s}}u)$ with $-2\pi uT_{\rm{s}}J_1(2\pi f_{\rm{D}}T_{\rm{s}}u)$ in $\M{\Sigma}(\V{s},\V{\theta})$, where $J_1(\cdot)$ is the Bessel function of the first kind.}

{Finally, by employing \eqref{eq:R9}, the \ac{da}-\ac{crlb} for \ac{mds} estimation in \ac{mimo} frequency-selective fading channel is obtained as
\begin{align}
\mathbb{V}{\rm{ar}}(\hat{f}_{\rm{D}})\geq {I}^{-1}(f_{\rm{D}})=\frac{1}{\frac{1}{2} {\rm{tr}} \Bigg{[}\bigg{(}\M{\Sigma}^{-1}(\V{s},\V{\theta})\frac{\partial \M{\Sigma}(\V{s},\V{\theta})}{\partial f_{\rm{D}}}\bigg{)}^2\Bigg{]}}.
\end{align}}
\newcounter{MYtempeqncnt}
\begin{figure*}[!htbp]
\normalsize
\setcounter{MYtempeqncnt}{\value{equation}}
\setcounter{equation}{18}
{
\begin{align}\label{eq:R14ioio}
\hspace{-0.2em}I(f_{\rm{D}})\hspace{-0.2em}=\hspace{-0.2em}{-\mathbb{E}\Bigg{\{}\frac{\partial^2\ln p({\RV{r}};\V{\varphi})}{\partial f_{\rm{D}}^2}\Bigg{\}}}
&\hspace{-0.2em}=\hspace{-0.2em}-\frac{1}{|M|^{N'n_{\rm{t}}}}\hspace{-0.2em}\int_{\RV{x}}\frac{\partial^2}{\partial f_{\rm{D}}^2  }
\Bigg{(}\ln \hspace{-0.2em} \sum_{i=1}^{\ |M|^{N'n_{\rm{t}}}}\frac{\exp \Big{(}-\frac{1}{2} {{{\RV{x}}^{\dag}}{\M{\Sigma}^{-1}(\V{c}_{\langle i \rangle},\V{\varphi})}{\RV{x}}} \Big{)}}{{ \det^{\frac{1}{2}} \big{(} \M{\Sigma}(\V{c}_{\langle i \rangle},\V{\varphi})\big{)}}}\Bigg{)}
\hspace{-0.6em}\sum_{q=1}^{\ |M|^{N'n_{\rm{t}}}}\hspace{-0.2em}\frac{\exp \Big{(}- \frac{1}{2} {{{\RV{x}}^{\dag}}{\M{\Sigma}^{-1}(\V{c}_{ \langle q \rangle},\V{\varphi})}{\RV{x}}} \Big{)}}{{(2\pi)^{Nn_{\rm{r}}} \det^{\frac{1}{2}} \big{(} \M{\Sigma}(\V{c}_{\langle q \rangle},\V{\varphi})\big{)}}}d{\RV{x}}.
\end{align}
\hrulefill}
\end{figure*}
\setcounter{equation}{12}
{\subsection{NDA-CRLB}
Let us consider that the symbols transmitted by each antenna are selected from a constellation with elements
$\{c_1 \ c_2 \ \cdots \ c_{|M|}\}$, where $\frac{1}{|M|}\sum_{i=1}^{|M|}|c_i|^2=1$.
The \ac{pdf} of the received vector $\RV{r}$ for NDA-MDS estimation is expressed as
\begin{align}\label{eq:R11w}
p({\RV{r}};\V{\varphi})&=\sum_{\RV{c}}p({\RV{r}},\RV{c};\V{\varphi}),
\end{align}
where $\RV{c}$ is the constellation vector as $\RV{c} \triangleq \big{[} {\RV{c}^{(1)}}^\dag \ {\RV{c}^{(2)}}^\dag   \ \cdots $ ${\RV{c}^{(n_{\rm{t}})}}^\dag \big{]}^\dag$,
$\RV{c}^{(m)} \triangleq \big{[}\bar{\rv{c}}_{1-L}^{(m)} \ \bar{\rv{c}}_{2-L}^{(m)} \ \cdots \ \bar{\rv{c}}_{N-1}^{(m)} \
\breve{\rv{c}}_{1-L}^{(m)} \ \breve{\rv{c}}_{2-L}^{(m)} \  \cdots \ $ $ \breve{\rv{c}}_{N-1}^{(m)} \big{]}^\dag$,
$\rv{c}_k^{(m)}=\bar {\rv{c}}_k^{(m)}+j\breve{\rv{c}}_k^{(m)}$
is the constellation point of the $m$th transmit antenna at time $k$, and $\V{\varphi}\triangleq [\V{\beta}^\dag \ \V{{\xi}}^\dag \ \V{{\vartheta}}^\dag \ f_{\rm{D}}]^\dag$ with $\V{\beta} \triangleq [\sigma_{\rm{s}_1}^2 \ \sigma_{\rm{s}_2}^2 \ \cdots  \sigma_{{\rm{s}}_{n_{\rm{t}}}}^2]^\dag$, and $\V{{\xi}}$ and $\V{{\vartheta}}$ are given in \eqref{uuuxazp}.}

{By employing the chain rule of probability and using $p(\RV{c}=\V{c}_{\langle i \rangle})=|M|^{-N'n_{\rm{t}}}$, $N'\triangleq N+L-1$,  one can write \eqref{eq:R11w} as
\begin{align}\label{eq:R11} \nonumber
p({\RV{r}};\V{\varphi})&=\sum_{\RV{c}}p({\RV{r}},\RV{c};\V{\varphi})=
\sum_{\V{c}}p(\RV{c}=\V{c}) p({\RV{r}}|{\RV{c}}={\V{c}};\V{\varphi}) \\
&=\frac{1}{|M|^{N'n_{\rm{t}}}}\sum_{i=1}^{\ |M|^{N'n_{\rm{t}}}}p({\RV{r}}|{\RV{c}}
={\V{c}_{\langle i \rangle}};\V{\varphi}),
\end{align}
where $\V{c}_{\langle i \rangle}$  represents the $i$th possible constellation vector at the transmit-side. }

{Similar to the \ac{da}-\ac{crlb}, $p({\RV{r}}|{\RV{c}}={\V{c}_{\langle i \rangle}};\V{\varphi})$ is Gaussian and
\begin{align}\label{eq:R12x}
p\big{(}{\RV{r}}|{\RV{c}}={\V{c}_{\langle i \rangle}};\V{\varphi}\big{)}=\frac{\exp \Big{(}-\frac{1}{2} {{{\RV{r}}^{\dag}}{\M{\Sigma}^{-1}(\V{c}_{\langle i \rangle},\V{\varphi})}{\RV{r}}} \Big{)}}{{(2\pi)^{Nn_{\rm{r}}} \det ^{\frac{1}{2}}  \big{(}\M{\Sigma}(\V{c}_{\langle i \rangle},\V{\varphi})\big{)}}},
\end{align}
where $\M{\Sigma} (\V{c}_{\langle i \rangle},\V{\varphi})\triangleq \mathbb{E} \big{\{}\RV{r}_{\langle i \rangle}\RV{r}_{\langle i \rangle}^\dag \big{\}}$ is the covariance matrix of the received vector
$\RV{r}_{\langle i \rangle}$ given the constellation vector is $\RV{c}=\V{c}_{\langle i \rangle}$, $i=1,2, \cdots , |M|^{N'n_{\rm{t}}}$. The $2Nn_{\rm{r}} \times 2Nn_{\rm{r}}$ covariance matrix $\M{\Sigma} (\V{c}_{\langle i \rangle},\V{\varphi})$ is
block diagonal as in \eqref{eq:covcov}, where its diagonal elements, i.e.,  $\M{\Sigma}_{\langle i \rangle}^{(n)} \triangleq \mathbb{E}\Big{\{}\RV{r}_{\langle i \rangle}^{(n)}{\RV{r}_{\langle i \rangle}^{(n)}}^\dag\Big{\}}$, $n \in \{1,2,\cdots,n_{\rm{r}}\}$,  are obtained as
\begin{subequations}
\begin{align}
\mathbb{E}  \Big{\{} \bar{\rv{r}}_{k,{\langle i \rangle}}^{(n)} &  \bar{\rv{r}}_{k+u,{\langle i \rangle}}^{(n)}\Big{\}}=\mathbb{E}  \Big{\{} \breve{\rv{r}}_{k,{\langle i \rangle}}^{(n)} \breve{\rv{r}}_{k,{\langle i \rangle}}^{(n)}\Big{\}} \\ \nonumber
&=\frac{1}{2}\sum\limits_{m = 1}^{{n_{\rm{t}}}} \sum\limits_{l = 1}^L {\sigma_{{{\rm{h}}_{(mn),l}}}^2\sigma_{{\rm{s}}_m}^2}\Big{(} \bar{{c}}_{k-l,{\langle i \rangle}}^{(m)}\bar{{c}}_{k+u-l,{\langle i \rangle}}^{(m)} \\ \nonumber
&\quad+\breve{{c}}_{k-l,{\langle i \rangle}}^{(m)}
\breve{{c}}_{k+u-l,{\langle i \rangle}}^{(m)}\Big{)}
 J_0(2\pi f_{\rm{D}}T_{\rm{s}}u)+\frac{\sigma_{{\rm{w}}_n}^2}{2} \delta_{u,0}
\end{align}
\begin{align}
\mathbb{E}  \Big{\{} \bar{\rv{r}}_{k,{\langle i \rangle}}^{(n)} &  \breve{\rv{r}}_{k+u,{\langle i \rangle}}^{(n)}\Big{\}}=-\mathbb{E}  \Big{\{} \breve{\rv{r}}_{k,{\langle i \rangle}}^{(n)} \bar{\rv{r}}_{k,{\langle i \rangle}}^{(n)}\Big{\}} \\ \nonumber
&=\frac{1}{2}\sum\limits_{m = 1}^{{n_{\rm{t}}}} \sum\limits_{l = 1}^L {\sigma_{{{\rm{h}}_{(mn),l}}}^2\sigma_{{\rm{s}}_m}^2}\Big{(} \bar{{c}}_{k-l,{\langle i \rangle}}^{(m)}\breve{{c}}_{k+u-l,{\langle i \rangle}}^{(m)} \\ \nonumber
&\quad-\breve{{c}}_{k-l,{\langle i \rangle}}^{(m)}
\bar{{c}}_{k+u-l,{\langle i \rangle}}^{(m)}\Big{)}
 J_0(2\pi f_{\rm{D}}T_{\rm{s}}u).
\end{align}
\end{subequations}}

{By substituting \eqref{eq:R12x} into \eqref{eq:R11}, one obtains
\begin{align}\label{eq:R12xdf}
p({\RV{r}};\V{\varphi})=\frac{1}{|M|^{N'n_{\rm{t}}}}\sum_{i=1}^{\ |M|^{N'n_{\rm{t}}}}\frac{\exp \Big{(}-\frac{1}{2} {{{\RV{r}}^{\dag}}{\M{\Sigma}^{-1}(\V{c}_{\langle i \rangle},\V{\varphi})}{\RV{r}}} \Big{)}}{{(2\pi)^{Nn_{\rm{r}}} \det ^{\frac{1}{2}} \big{(} \M{\Sigma}(\V{c}_{\langle i \rangle},\V{\varphi})\big{)}}}.
\end{align}}
{Finally, by employing \eqref{eq:R12xdf}, the \ac{nda}-\ac{crlb} for \ac{mds} estimation in \ac{mimo} frequency-selective fading channel is expressed as
\begin{align}\label{eq:R13}
\mathbb{V}{\rm{ar}}(\hat{f}_{\rm{D}})\geq {I}^{-1}(f_{\rm{D}})=\frac{1}{-\mathbb{E}\Big{\{}\frac{\partial^2\ln p({\RV{r}};\V{\varphi})}{\partial f_{\rm{D}}^2}\Big{\}}},
\end{align}
where ${I}(f_{\rm{D}})$ is given in \eqref{eq:R14ioio} on the top of this page, and $\int_{\RV{x}} \triangleq \int_{x_1}\int_{x_2}\cdots\int_{x_{(2 N n_{\rm{r}})}}$.
As seen, there is no an explicit expression for \eqref{eq:R14ioio}, and thus, for the \ac{crlb} in \eqref{eq:R13}. Therefore, numerical methods are used to solve \eqref{eq:R14ioio} and \eqref{eq:R13}.}

\begin{figure*}[!t]
\normalsize
\setcounter{MYtempeqncnt}{\value{equation}}
\setcounter{equation}{27}
\begin{align}\label{T44} \nonumber
\kappa_u^{(n)} &= {\mathbb E}\left\{\big{|}\rv{r}_k^{(n)}{\big{|}^2}|\rv{r}_{k + u}^{(n)}{\big{|}^2}\right\} = \sum\limits_{m = 1}^{{n_{\rm{t}}}} {\sum\limits_{l = 1}^L {\,\mathbb{E}\left\{\big{|}\rv{h}_{k,l}^{(mn)}{\big{|}^2}\big{|}\rv{h}_{k + u,l}^{(mn)}{\big{|}^2}\right\}} } \,\,\sigma _{{{\rm{s}}_m}}^4 + \sum\limits_{{m_1} = 1}^{{n_{\rm{t}}}} {\sum\limits_{{m_2} \ne {m_1}}^{n_{\rm{t}}}{\sum\limits_{{l} = 1}^L {\,\mathbb{E}\left\{\big{|}\rv{h}_{k,{l}}^{({m_1n})}{\big{|}^2}\big{|}\rv{h}_{k + u,{l}}^{({m_2n})}{\big{|}^2}\right\}} } } \,\,\sigma _{{\rm{s}}_{m_1}}^2\sigma _{{\rm{s}}_{m_2}}^2\\ \nonumber
& \quad + \,\sum\limits_{m = 1}^{{n_{\rm{t}}}} {\sum\limits_{{l_1} = 1}^L {\sum\limits_{{l_2} \ne {l_1}}^L {\mathbb {E}\left\{\big{|}\rv{h}_{k,{l_1}}^{(mn)}{\big{|}^2}\big{|}\rv{h}_{k + u,{l_2}}^{(mn)}{\big{|}^2}\right\}} } } \,\sigma _{{{\rm{s}}_m}}^4 + \sum\limits_{{m_1} = 1}^{{n_{\rm{t}}}} {\sum\limits_{{m_2} \ne {m_1}}^{{n_{\rm{t}}}} {\sum\limits_{{l_1} = 1}^L {\sum\limits_{{l_2} \ne {l_1}}^L {\mathbb{ E}\left\{\big{|}\rv{h}_{k,{l_1}}^{({m_1n})}{\big{|}^2}\big{|}\rv{h}_{k + u,{l_2}}^{({m_2n})}{\big{|}^2}\right\}\,\sigma _{{\rm{s}}_{m_1}}^2\sigma _{{\rm{s}}_{m_2}}^2} } } }  \\
& \quad+ 2\sigma _{{\rm{w}}_n}^2\sum\limits_{m = 1}^{{n_{\rm{t}}}} {\sum\limits_{l = 1}^L {\mathbb {E}\left\{\big{|}\rv{h}_{k,l}^{(mn)}{\big{|}^2}\right\}\,} } \sigma _{{{\rm{s}}_m}}^2\, + \sigma _{{\rm{w}}_n}^4,\,\,\,\,\,\,\,\,\,\,\,\,\,\,u \ge L.
\end{align}
\hrulefill
\vspace*{4pt}
\end{figure*}

{\section{ML estimation for MDS}\label{section:ML}
In this section, we derive the \ac{da}- and \ac{nda}-\ac{mle}s for \ac{mds} in MIMO frequency-selective fading channel.}
\setcounter{equation}{19}
{\subsection{\ac{da}-\ac{mle} for \ac{mds}}
The \ac{da}-\ac{mle} for $f_{\rm{D}}$ is
obtained as
\begin{align}
\hat{f}_{\rm{D}}=\underset{{f}_{\rm{D}}} {\arg \max} \ p({\RV{r}}|\V{s};\V{\theta}),
\end{align}
where $p({\RV{r}}|\V{s};\V{\theta})$ is given in \eqref{eq:R5}. Since $p({\RV{r}}|\V{s};\V{\theta})$ is a differentiable function, the \ac{da}-\ac{mle} for $f_{\rm{D}}$ is obtained from
\begin{align}\label{eq:MLDA}
\frac{\partial \ln p({\RV{r}}|{\V{s}};\V{\theta})}{\partial f_{\rm{D}}}=0.
\end{align}
By substituting \eqref{eq:R5} into \eqref{eq:MLDA} and after some mathematical manipulations, one obtains
\begin{align}\label{eq:MLDA2}
\frac{\partial \ln p({\RV{r}}|{\V{s}};\V{\theta})}{\partial f_{\rm{D}}}&=-\frac{1}{2} \text{tr} \Bigg{[}\M{\Sigma}^{-1}({\V{s}},\V{\theta})\frac{\partial \M{\Sigma}({\V{s}},\V{\theta})}{\partial f_{\rm{D}}}\Bigg{]}
\\ \nonumber
&\quad+\frac{1}{2}\RV{r}^\dag \M{\Sigma}^{-1}({\V{s}},\V{\theta})\frac{\partial \M{\Sigma}({\V{s}},\V{\theta})}{\partial f_{\rm{D}}}\M{\Sigma}^{-1}({\V{s}},\V{\theta})\RV{r}.
\end{align}
As seen in \eqref{eq:MLDA2}, there is no closed-form solution for \eqref{eq:MLDA}. Thus, numerical methods need to be used to obtain solution. By employing the Fisher-scoring method \cite{longford1987fast},\footnote{{The Fisher-scoring method replaces the Hessian matrix in
the Newtown-Raphson method with the negative of the Fisher information matrix \blue{\cite{kay1993fundamentals}}.}}
the solution of \eqref{eq:MLDA2} can be iteratively obtained as
\begin{align}\label{eq:new}
\hat{f}_{\rm{D}}^{[t+1]}=\hat{f}_{\rm{D}}^{[t]}+{I}^{-1}(f_{\rm{D}})\frac{\partial \ln p({\RV{r}}|{\V{s}};\V{\theta})}{\partial f_{\rm{D}}}\Bigg{|}_{{f}_{\rm{D}}=\hat{f}_{\rm{D}}^{[t]}},
\end{align}
where ${I}(f_{\rm{D}})$ and $\frac{\partial \ln p({\RV{r}}|{\V{s}};\V{\theta})}{\partial f_{\rm{D}}}$ are given in
\eqref{eq:R9} and \eqref{eq:MLDA2}, respectively.}
{\subsection{\ac{nda}-\ac{mle} for MDS}
Similar to the \ac{da}-\ac{mle}, the \ac{nda}-\ac{mle} for \ac{mds} is obtained from
\begin{align}
\hat{f}_{\rm{D}}=\underset{{f}_{\rm{D}}} {\arg \max} \ p({\RV{r}};\V{\varphi}),
\end{align}
where $p({\RV{r}};\V{\varphi})$ is given in \eqref{eq:R12xdf}.
Since $p({\RV{r}};\V{\varphi})$ is a linear combination of differentiable functions, the \ac{nda}-\ac{mle} for $f_{\rm{D}}$ is obtained from
\begin{align}\label{eq:MLNDA}
\frac{\partial  \ln  p({\RV{r}};\V{\varphi})}{\partial f_{\rm{D}}}=0.
\end{align}
By substituting \eqref{eq:R12xdf} into \eqref{eq:MLNDA} and after some algebra, one obtains
\begin{align}\label{eq:MLNDA2}\nonumber
&\sum_{i=1}^{\ |M|^{N'n_{\rm{t}}}}\Bigg{\{}
\frac{{{{\RV{r}}^{\dag}}{\M{\Sigma}^{-1}(\V{c}^{ \langle i \rangle},\V{\varphi})}\frac{\partial {\M{\Sigma}(\V{c}_{\langle i \rangle},\V{\varphi})}}{\partial f_{\rm{D}}}{\M{\Sigma}^{-1}(\V{c}_{ \langle i \rangle},\V{\varphi})}{\RV{r}}}}{{{\det}^{\frac{1}{2}}  \M{\Sigma}(\V{c}_{\langle i \rangle},\V{\varphi})}} \\
&\quad \,\,\,\,\,\,\,\,\,\,\,\,\,\,\,\,\,\ -\frac{
\text{tr}\Big{[}\M{\Sigma}^{-1}(\V{c}_{ \langle i \rangle},\V{\varphi})\frac{\partial \M{\Sigma}(\V{c}_{\langle i \rangle},\V{\varphi})}{\partial f_{\rm{D}}} \Big{]}}{{\det}^{\frac{1}{2}}  \M{\Sigma}(\V{c}_{\langle i \rangle},\V{\varphi})}\Bigg{\}}=0
\end{align}
Similar to the \ac{da}-\ac{mle}, there is no closed-form solution for \eqref{eq:MLNDA2}; thus, numerical methods are used to solve \eqref{eq:MLNDA2}.}

\section{\ac{nda}-\ac{mb} estimation of \ac{mds}}\label{section:mb}
In this section, we propose an \ac{nda}-\ac{mb} \ac{mds} estimator for \ac{miso} systems under frequency-selective Rayleigh fading channel by employing the fourth-order moment of the received signal. Then, an extension of the proposed estimator to the MIMO systems is provided.

\subsection{\ac{nda}-\ac{mbe} for \ac{mds} in \ac{miso} Systems}\label{section:mba}
{Let us assume that the parameter vector $\V{\varphi}= [\V{\beta}^\dag \ \V{{\xi}}^\dag $ $ \V{{\vartheta}}^\dag \ f_{\rm{D}}]^\dag$ is unknown at the receive-side. The statistical \ac{mb} approach enables us to propose an \ac{nda}-\ac{mbe} to estimate $f_{\rm{D}}$ without any priori knowledge of $\V{\beta}$, $\V{\xi}$, and $\V{\vartheta}$.}
Let us consider the fourth-order two-conjugate moment of the received signal at the $n$th receive antenna, defined as
%\footnote{As the MISO scenario is considered in this section, the index $n$ of the receive antenna is dropped for $\rv{r}$.}
\begin{equation}\label{eq:TT3}
{\kappa_u^{(n)}} \buildrel \Delta \over =  {\mathbb E}\left\{ {\big{|}\rv{r}_k^{(n)}{\big{|}^2}\big{|}\rv{r}_{k +u}^{(n)}{\big{|}^2}}\right\}.
\end{equation}
With the transmitted symbols, $\rv{s}_k^{(m)}$, $m=1,...,n_{\rm{t}}$ being independent, drawn from symmetric complex-valued constellation points,\footnote{$\mathbb{E}\big{\{}{(\rv{s}_k^{(m)})^2}\big{\}} = 0$ for $M$-ary phase-shift-keying (PSK) and quadrature amplitude modulation (QAM), $M>2$ \cite{swami2000hierarchical}.} and with $u \ge L$, $\kappa_u^{(n)}$ is expressed as in \eqref{T44} at the top of this page (see Appendix \ref{apd:proofI} for proof).

\setcounter{equation}{28}

{By employing the first-order autoregressive model of the Rayleigh fading channel, one can write \cite{baddour2005autoregressive,sadeghi2008finite}
\begin{align}\label{eq:1auto}
{\rv{h}_{k,l}^{(mn)}} = {\Psi _u}{\rv{h}_{k+u,l}^{(mn)}}+ {\rv{v}_{k,l}^{(mn)}},
\end{align}
where ${\Psi _u} \triangleq J_0(2\pi f_{\rm{D}} T_{\rm{s}}u)$ and ${\rv{v}_{k,l}^{(mn)}}$ is a zero-mean complex-valued Gaussian white process with variance ${\mathbb E}\{|{\rv{v}_{k,l}^{(mn)}}|^2\} = (1 - |{\Psi _u}{|^2})\sigma _{{\rm{h}}_{(mn),l}}^2$, which is independent of ${\rv{h}_{k,l}^{(mn)}}$.}

{By using \eqref{eq:1auto} and exploiting the property of a complex-valued Gaussian random variable $\rv{z} \sim {\cal{N}_{\rm{c}}}\left( {0,\sigma _{\rm{z}}^2} \right)$ that ${ {\mathbb E}}\{|\rv{z}{|^{2n}}\} = n!\sigma _{\rm{z}}^{2n}$ \cite{reed1962moment}, one obtains}
\begin{align}\label{eq:TT7}
&{\mathbb E}\left\{\big{|}\rv{h}_{k,l}^{(mn)}{\big{|}^2}|\rv{h}_{k + u,l}^{(mn)}{\big{|}^2}\right\} \\[1.5ex]    \nonumber
 &\quad= {\mathbb E}\bigg{\{}\Big{|}({J_0}(2\pi {f_{\rm{d}}T_s}u)\rv{h}_{k + u,l}^{(mn)} + {\rv{v}_{k,l}^{(mn)}}){\Big{|}^2}\big{|}\rv{h}_{k + u,l}^{(mn)}{\big{|}^2}\bigg{\}}\\[1.5ex] \nonumber
 & \quad={J_0^2}(2\pi{f_{\rm{d}}T_{\rm{s}}}u){\mathbb E}\Big{\{}\big{|}\rv{h}_{k + u,l}^{(mn)}{\big{|}^4}\Big{\}}
+ {\mathbb E}\left\{\big{|}{\rv{v}_{k,l}^{(mn)}}{\big{|}^2}|\rv{h}_{k + u,l}^{(mn)}{\big{|}^2}\right\}\\[1.5ex] \nonumber
& \quad \quad + {J_0}(2\pi {f_{\rm{d}}T_s}u){\mathbb E}\left\{\rv{h}_{k + u,l}^{(mn)}\big{|}\rv{h}_{k + u,l}^{(mn)}{\big{|}^2}{({\rv{v}_{k,l}^{(mn)}})^*}\right\}\\[1.5ex] \nonumber
&\quad \quad + {J_0}(2\pi {f_{\rm{d}}T_{\rm{s}}}u){\mathbb E}\left\{{(\rv{h}_{k + u,l}^{(mn)})^*}\big{|}\rv{h}_{k + u,l}^{(mn)}{\big{|}^2}({\rv{v}_{k,l}^{(mn)}})\right\} \\[1.5ex] \nonumber
&\quad =2{J_0^2}(2\pi {f_{\rm{d}}T_{\rm{s}}}u)\sigma _{{\rm{h}}_{(mn),l}}^4 + \big{(}1 - {J_0^2}(2\pi {f_{\rm{d}}T_{\rm{s}}}u)\big{)}\sigma _{{\rm{h}}_{(mn),l}}^4 \\[1.5ex] \nonumber
& \quad=\big{(}1 + {J_0^2}(2\pi {f_{\rm{d}}T_{\rm{s}}}u)\big{)}\sigma _{{\rm{h}}_{(mn),l}}^4\,\,\,\,\,m=1,...n_{\rm{t}},\,\,\, l=1,...,L.
\end{align}

With the channel {taps} $l_1$ and $l_2$ being uncorrelated for each transmit antenna, i.e., $\mathbb{E}\big{\{}\rv{h}_{k,l_1}^{(mn)}{(\rv{h}_{k,l_2}^{(mn)})^*}\big{\}}=\sigma_{{\rm{h}}_{(mn),l_1}}^2\delta_{l_1 , l_2}$ and employing
\begin{align}\label{set}
\mathbb{E}&\left\{\big{|}\rv{h}_{k,l_1}^{(m_1n)}\big{|}^2\big{|}{\rv{h}_{k+u,l_2}^{(m_2n)}\big{|}^2}\right\} \\ \nonumber
&=\sigma_{{\rm{h}}_{(m_1n),l_1}}^2 \sigma_{{\rm{h}}_{(m_2n),l_2}}^2\Big{[}(1-\delta_{l_1 , l_2})(1-\delta_{m_1 , m_2}) \\ \nonumber
& \quad \quad +\delta_{l_1,l_2}(1-\delta_{m_1 , m_2})
  +(1-\delta_{l_1 , l_2})\delta_{m_1 , m_2}\Big{]} \\ \nonumber
& \quad \quad  + \sigma_{{\rm{h}}_{(m_1n),l_1}}^4 \big{(}1 + {J_0^2}(2\pi {f_{\rm{d}}T_{\rm{s}}}u)\big{)}
\delta_{m_1,m_2}\delta_{l_1 , l_2},
\end{align}
one can write \eqref{T44} as
\begin{align}\label{eq:TT8}
\kappa_u^{(n)} &= \sum\limits_{m = 1}^{{n_{\rm{t}}}} {\sum\limits_{l = 1}^L {\sigma _{{{\rm{h}}_{(mn),l}}}^4} } \sigma _{{\rm{s}}_{{m}}}^4\big{(}1 + {J_0^2}({2\pi {f_{\rm{d}}}{T_{\rm{s}}}u})\big{)} \\ \nonumber
& \quad+ \sum\limits_{{m_1} = 1}^{{n_{\rm{t}}}} {\sum\limits_{{m_2} \ne {m_1}}^{n_{\rm{t}}} {\sum\limits_{{l} = 1}^L {\sigma _{{\rm{h}}_{(m_1n),l}}^2}{\sigma _{{\rm{h}}_{(m_2n),l}}^2} } } \sigma _{{\rm{s}}_{m_1}}^2\sigma _{{\rm{s}}_{m_2}}^2 \\ \nonumber
& \quad + \sum\limits_{m = 1}^{{n_{\rm{t}}}} {\sum\limits_{{l_1} = 1}^L {\sum\limits_{{l_2} \ne {l_1}}^L {\sigma _{{\rm{h}}_{(mn),l_1}}^2} \sigma _{{\rm{h}}_{(mn),l_2}}^2} } \sigma _{{{\rm{s}}_m}}^4\\ \nonumber
& \quad + \sum\limits_{{m_1} = 1}^{{n_{\rm{t}}}} {\sum\limits_{{m_2} \ne {m_1}}^{{n_{\rm{t}}}} {\sum\limits_{{l_1} = 1}^L {\sum\limits_{{l_2} \ne {l_1}}^L {\sigma _{{\rm{h}}_{(m_1n),l_1}}^2\sigma _{{\rm{h}}_{(m_2n),l_2}}^2\sigma _{{\rm{s}}_{m_1}}^2\sigma _{{\rm{s}}_{m_2}}^2} } } } \\ \nonumber
& \quad + 2\sigma _{{\rm{w}}_n}^2\sum\limits_{m = 1}^{{n_{\rm{t}}}} {\sum\limits_{l = 1}^L {\sigma _{{\rm{h}}_{(mn),l}}^2} } \sigma _{{\rm{s}}_{{m}}}^2 +\sigma _{{\rm{w}}_n}^4.
\end{align}

\begin{figure*}[!h]
\normalsize
\setcounter{MYtempeqncnt}{\value{equation}}
\setcounter{equation}{41}
{
\begin{align}\label{eq:r450o}
\hat {\rv{f}}_{\rm{D}}^{[t+1]}=\hat {\rv{f}}_{\rm{D}}^{[t]}-\frac{\sum\limits_{u = {U_{\min}}}^{{M_{\max }}}{{{8\pi T_{\rm{s}} u \left( {{{\hat {\rv{\Psi}} }_u^{(n)}} - J_0^2(2\pi {\rv{f}_{\rm{D}}^{[t]}}{T_{\rm{s}}}u)} \right)}}}
J_0\big{(}2\pi \rv{f}_{\rm{D}}^{[t]}{T_{\rm{s}}}u\big{)}J_1\big{(}2\pi \rv{f}_{\rm{D}}^{[t]}{T_{\rm{s}}}u\big{)}}{\frac{\partial^2}{\partial f_{\rm{D}}^2}\sum\limits_{u = {U_{\min}}}^{{U_{\max }}} {{{\left( {{\hat {\rv{\Psi}}_u^{(n)}} - J_0^2(2\pi {f_{\rm{D}}}{T_{\rm{s}}}u)} \right)}^2}_{\big{|}f_{\rm{D}}=\rv{f}_{\rm{D}}^{[t]}}}
}
\end{align}}
{
\begin{align} \nonumber
& {\frac{\partial^2}{\partial f_{\rm{D}}^2}\sum\limits_{u = {U_{\min}}}^{{U_{\max }}} {{{\left( {{\hat {\rv{\Psi}}_u} - J_0^2(2\pi {f_{\rm{D}}}{T_{\rm{s}}}u)} \right)}^2}_{\big{|}f_{\rm{D}}=\rv{f}_{\rm{D}}^{[t]}}}
}=\sum\limits_{u = {U_{\min}}}^{{M_{\max }}}\Bigg{\{}32\pi^2T_{\rm{s}}^2u^2J_0^2\big{(}2\pi \rv{f}_{\rm{D}}^{[t]}{T_{\rm{s}}}u\big{)}J_1^2\big{(}2\pi \rv{f}_{\rm{D}}^{[t]}{T_{\rm{s}}}u\big{)} \\ \nonumber
&\quad \quad \quad+8\pi T_{\rm{s}}u\Bigg{(}2\pi T_{\rm{s}}u\bigg{(}J_0^2\big{(}2\pi \rv{f}_{\rm{D}}^{[t]}{T_{\rm{s}}}u\big{)}-J_1^2\big{(}2\pi \rv{f}_{\rm{D}}^{[t]}{T_{\rm{s}}}u\big{)}\bigg{)}-\frac{J_0\big{(}2\pi \rv{f}_{\rm{D}}^{[t]}{T_{\rm{s}}}u\big{)}J_1\big{(}2\pi \rv{f}_{\rm{D}}^{[t]}{T_{\rm{s}}}u\big{)}}{\rv{f}_{\rm{D}}^{[t]}} \Bigg{)}\left( {{{\hat {\rv{\Psi}} }_{u}^{(n)}} - J_0^2(2\pi {\rv{f}_{\rm{D}}^{[t]}}{T_{\rm{s}}}u)} \right)\Bigg{\}}
\end{align}}
\hrulefill
\vspace*{3pt}
\setcounter{equation}{32}
\end{figure*}

Further, let us consider the second-order moment of the received signal, i.e., ${\mu_2^{(n)}} \buildrel \Delta \over = \mathbb{E}\{|{\rv{r}_k^{(n)}}{|^2}\}$. By using \eqref{eq:TT1}, it can be easily shown that
\begin{equation}\label{eq:TT9}
\mu_2^{(n)}= \sum\limits_{m = 1}^{{n_{\rm{t}}}} {\sum\limits_{l = 1}^L {\sigma _{{\rm{h}}_{(mn),l}}^2} } \sigma _{{\rm{s}}_{{m}}}^2 + \sigma _{{\rm{w}}_n}^2.
\end{equation}
By employing \eqref{eq:TT8} and \eqref{eq:TT9}, one obtains the normalized squared \ac{af} of the fading channel as {(see Appendix \ref{ap2xc} for proof)}
\begin{equation}\label{eq:TT11y}
\Psi_u \buildrel \Delta \over = J_0^2(2\pi {f_{\rm{d}}}{T_{\rm{s}}}u) = \eta^{(n)} \bigg{(}{{\kappa_{u}^{(n)}} - \Big{(}\mu_2^{(n)}\Big{)}^2}\bigg{)},
\end{equation}
where
$\eta^{(n)} ={1 \mathord{\left/
 {\vphantom {1 {}}} \right.
 \kern-\nulldelimiterspace} {\sum\limits_{m = 1}^{{n_{\rm{t}}}} {\sum\limits_{l = 1}^L {\sigma _{{\rm{h}}_{(mn),l}}^4} } \sigma _{{\rm{s}}_{{m}}}^4}}$.

For non-constant modulus constellations, $\eta^{(n)}$ is
expressed in terms of $\mu_4^{(n)} \buildrel \Delta \over = \mathbb{E}\big{\{}|{\rv{r}_k^{(n)}}{|^4}\big{\}}$ and $\mu_2^{(n)}$ as (see Appendix \ref{ap3} for proof)
\begin{equation}\label{eq:TT12x}
\eta^{(n)}  = \frac{{2({\Omega _{\rm{s}}} - 1)}}{{{\mu_4^{(n)}} - 2\Big{(}\mu_2^{(n)}\Big{)}^2}}\,,
\end{equation}
where $\Omega_{\rm{s}}=\frac{1}{|M|}\sum_{i=1}^{|M|}|c_i|^4$  is a constant, and $1<\Omega_{\rm{s}}\le2$.\footnote{For 16-QAM, 64-QAM, and complex-valued zero-mean Gaussian signals, $\Omega_s$ is 1.32, 1.38, and 2, respectively \cite{swami2000hierarchical}.}

Finally, substituting \eqref{eq:TT12x} into \eqref{eq:TT11y} yields
\begin{equation}\label{eq:TT13}
\Psi_u \buildrel \Delta \over = 2({\Omega _{\rm{s}}} - 1)\frac{{{\kappa_u^{(n)}} - \Big{(}\mu_{2}^{(n)}\Big{)}^2}}{{{\mu_{4}^{(n)}} - 2\Big{(}\mu_2^{(n)}\Big{)}^2}}.
\end{equation}

As seen, the normalized squared \ac{af} of the fading channel is expressed as a non-linear function of the $\mu_2^{(n)}$, $\mu_4^{(n)}$, and $\kappa_u^{(n)}$.
In practice, statistical moments are estimated by time
averages of the received signal.
For \eqref{eq:TT13}, the following
estimators of the moments are employed
\begin{align} \label{eq:TT14}
{{\hat {\rv{\mu}} }_2^{(n)}} &= \frac{1}{N}\sum\limits_{k = 1}^N {\big{|}{\rv{r}_k^{(n)}}{\big{|}^2}} \\ \nonumber
\hat{\rv{\mu}}_{4}^{(n)} &= \frac{1}{N}\sum\limits_{k = 1}^N {|\rv{r}_k^{(n)}{\big{|}^4}}\\ \nonumber
\hat{\rv{\kappa}}_{u}^{(n)} &= \frac{1}{N-u}\sum\limits_{k = 1}^{N-u} {\big{|}\rv{r}_k^{(n)}{\big{|}^2}\big{|}\rv{r}_{k +u}^{(n)}{\big{|}^2}},
\end{align}
where  $u\geq L>0$.

By substituting the corresponding estimators in \eqref{eq:TT13}, the estimate of the normalized squared AF is
given as
\begin{equation}\label{eq:TT19}
\hat {\rv{\Psi}}_u^{(n)} \buildrel \Delta \over = 2({\Omega_{\rm{s}}} - 1)\frac{{{\hat {\rv{\kappa}}_{u}^{(n)}} - \Big{(}\hat{ \rv{\mu}}_{2}^{(n)}\Big{)}^2}}{{{\hat{\rv{\mu}}_{4}^{(n)}} - 2\Big{(}\hat{\rv{\mu}}_2^{(n)}\Big{)}^2}}.
\end{equation}

Now, based on \eqref{eq:TT11y} and \eqref{eq:TT19}, the problem of \ac{mds} estimation can be formulated as a non-linear regression problem. Given the estimated normalized squared AF, $\hat{\rv{\Psi}}_u^{(n)}$, the  non-linear regression model assumes that the relationship between ${\hat{\rv{\Psi}}}_u^{(n)}$ and ${{{\Psi}}}_u$ is  modeled through a disturbance term or error variable ${\rv{\epsilon}}_u^{(n)}$ as \cite{gallant2009nonlinear,draper1966applied}
\begin{align}\label{eq:TT23}
\hat{\rv{\Psi}}_u^{(n)}&={\Psi}_u+{\rv{\epsilon}}_u^{(n)} \\ \nonumber
&=J_0^2(2\pi f_{\rm{D}}T_{\rm{s}}u)+{\rv{\epsilon}}_u^{(n)}, \,\,\,\,\,\ u=U_{\min},\hdots,U_{\max},
 \end{align}
where $U_{\rm{min}}$ and $U_{\rm{max}}$ are the maximum and minimum delay lags, respectively.

To solve the non-linear regression problem in \eqref{eq:TT23}, the LS curve-fitting optimization technique is employed.
Based on the LS curve-fitting optimization, the estimate of $f_{\rm{D}}$, i.e., $\hat {\rv{f}}_{\rm{D}}$, is obtained through minimizing the sum of the squared residuals (SSR) as \cite{draper1966applied}
\begin{equation}\label{eq:TT20}
\begin{aligned}
& \underset{f_{\rm{D}}}{\text{minimize}}
& & \sum\limits_{u = {U_{\min}}}^{{U_{\max }}} {{{\left( {{\hat {\rv{\Psi}}_u^{(n)}} - J_0^2(2\pi {f_{\rm{D}}}{T_{\rm{s}}}u)} \right)}^2}} \\
& \text{subject to}
& & f_{\rm{l}}\leq f_{\rm{D}} \leq f_{\rm{h}},
\end{aligned}
\end{equation}
where $f_{\rm{l}}$ and $f_{\rm{h}}$ are the minimum and maximum possible \ac{mds}s, respectively.
To obtain $\hat {\rv{f}}_{\rm{D}}$, we consider the
derivative of the SSR with respect to $f_{\rm{D}}$
and set it equal to zero {as follows}:
\begin{align}\label{eq:16x}
\sum\limits_{u = {U_{\min}}}^{{M_{\max }}}&{{{8\pi T_{\rm{s}} u \left( {{{\hat {\rv{\Psi}} }_{u}^{(n)}} - J_0^2(2\pi {f_{\rm{D}}}{T_{\rm{s}}}u)} \right)}}} \\ \nonumber
&J_0(2\pi f_{\rm{D}}{T_{\rm{s}}}u)J_1(2\pi f_{\rm{D}}{T_{\rm{s}}}u)=0.
\setcounter{equation}{42}
\end{align}
As seen, for the non-linear regression, the derivative in \eqref{eq:16x}
is a function of $f_{\rm{D}}$. Thus, an explicit solution for $\hat {\rv{f}}_{\rm{D}}$
cannot be obtained. However, numerical methods
\cite{rmm} can be employed to solve the LS curve-fitting optimization problem in \eqref{eq:TT20}.

{By employing the Newton-Raphson method, $\hat {\rv{f}}_{\rm{D}}$ can be iteratively obtained as it is shown in \eqref{eq:r450o} at the top of next page.
The main problem with
the Newton-Raphson method is that it suffers from the convergence problem \cite{kay1993fundamentals}.}
Since the parameter space for the \ac{mds} estimation is one-dimensional, {the grid search method can be employed, which ensures the global optimality of the solution.}
With the grid search method, the parameter space, i.e., $[f_{\rm{l}} , f_{\rm{h}}]$ is discretized as a grid with step size $\delta$, and the value which minimizes SSR is considered as the estimated $f_{\rm{D}}$. This procedure can be performed in two steps, including a rough estimate of the \ac{mds}, $\hat {\rv{F}}_{\rm{D}}^{(\rm{r})}$, by choosing a larger step size $\Delta$ followed by a fine estimate, $\hat {\rv{F}}_{\rm{D}}^{(\rm{s})}$, through small grid step size $\delta$ around the rough estimate, i.e., $\big{[}\hat {\rv{F}}_{\rm{D}}^{(\rm{r})}-\Delta,\hat {\rv{F}}_{\rm{D}}^{(\rm{r})}+\Delta\big{]}$.
A formal description
of the proposed  \ac{nda}-\ac{mbe} for \ac{mds} in \ac{miso} frequency-selective channel is presented in Algorithm \ref{al:1}.

It is worth noting that $f_{\rm{D}}$ can be estimated by using a downsampled version of  ${{\hat {\rv{\Psi}} }_u^{(n)}}$. For the case of uniform downsampling, i.e., $u=\ell u_{\rm{s}}$, the SSR is given as
\begin{equation}
\sum\limits_{\ell=0}^{N_{\rm{la}}-1}{{{\left( {{\hat {\rv{\Psi}}}_{U_{\rm{min}}+\ell u_{\rm{s}}}^{(n)} - { {{\Psi}} }_{ U_{\rm{min}}+\ell u_{\rm{s}}}} \right)}^2}} ,
\end{equation}
where $u_{\rm{s}}$ is the downsampling period expressed in delay lags, $N_{\rm{la}}$ is the number of delay lag,
\begin{equation}\label{eq:TTXX}
{{{\Psi}}_{\ell u_{\rm{s}}}} = J_0^2\Big{(}2\pi {f_{\rm{D}}}{T_{\rm{s}}}(U_{\rm{min}}+\ell u_{\rm{s}})\Big{)},
\end{equation}
and
\begin{equation}\label{eq:TTYY}
{\hat {\rv{\Psi}}_{U_{\rm{min}}+\ell u_{\rm{s}}}}^{(n)} = 2({\Omega _s} - 1)\frac{{{\hat {\rv{\kappa}}_{U_{\rm{min}}+\ell u_{\rm{s}}}^{(n)}} - \Big{(}\hat{ \rv{\mu}}_{2}^{(n)}\Big{)}^2}}{{{\hat{\rv{\mu}}_{4}^{(n)}} - 2\Big{(}\hat{\rv{\mu}}_2^{(n)}\Big{)}^2}}.
\end{equation}
The downsampled version of ${{\hat {\rv{\Psi}} }_u^{(n)}}$ is usually employed for the rough \ac{mds} estimation, where $\Delta$ is a large value.
  \begin{algorithm}[!t]
    \caption{:  \ac{nda}-\ac{mbe} for \ac{mds} in MISO systems}\label{euclid}
    \begin{algorithmic}[1]
\State {Set $f_{\rm{l}}$, $f_{\rm{h}}$, $\Delta$, and $\delta$ }
\State {Acquire the measurements $\big{\{} {{\rv{r}_k^{(n)}}} \big{\}}_{k = 1}^N$}
\State {Estimate the statistics $\hat {\rv{\mu}}_2^{(n)}$, $\hat {\rv{\mu}}_4^{(n)}$, and $\hat {\rv{\kappa}}_u^{(n)}$, by employing \eqref{eq:TT14}}
\State {Compute ${{\hat {\rv{\Psi}} }_u^{(n)}}$, $\forall u \in \big{\{}U_{\rm{min}},\hdots,U_{\rm{max}}\big{\}}$ by using \eqref{eq:TT19}}
\State {Obtain $\hat {\rv{F}}_{\rm{D}}^{(\rm{r})}$ by solving the minimization problem in \eqref{eq:TT20} through  the grid search method with grid step size $\Delta$}
\State {Obtain $\hat {\rv{F}}_{\rm{D}}^{(\rm{s})}$ by solving the minimization problem in \eqref{eq:TT20} through  the grid search method over $\big{[}\hat {\rv{F}}_{\rm{D}}^{(\rm{r})}-\Delta,\hat {\rv{F}}_{\rm{D}}^{(\rm{r})}+\Delta\big{]}$ with grid step size $\delta$}
\State {$\hat{\rv{F}}_{\rm{D}}=\hat{\rv{F}}_{\rm{D}}^{(\rm{s})}$}
    \end{algorithmic}\label{al:1}
  \end{algorithm}

\subsection{\ac{nda}-\ac{mbe} for \ac{mds} in \ac{mimo} Systems}\label{section:mbb}
The performance of the proposed \ac{nda}-\ac{mbe} for \ac{mds} in MISO system can be improved when employing multiple receive antennas due to the spatial diversity, by combining the estimated normalized squared AFs, ${\hat {\rv{\Psi}}_u^{(n)}}$, $n=1,...,n_{\rm{r}}$ as
{\begin{align}\label{eq:weight}
{\widetilde{{\rv{\Psi}}}}_u = \sum\limits_{n = 1}^{{n_{\rm{r}}}} \lambda_{u}^{(n)} {\hat {\rv{\Psi}}_u^{(n)}},
\end{align}
where $\V{\Lambda}_{u} \triangleq \big{[}\lambda_{u}^{(1)} \ \lambda_{u}^{(2)} \ \cdots\ \lambda_{u}^{(n_{\rm{r}})}\big{]}^\dag $, with $\sum_{n=1}^{n_{\rm{r}}}\lambda_{u}^{(n)}=1$,  is the weighting vector.
Let us define $\hat {\RV{\Psi}}_u \triangleq \big{[}{\hat {\rv{\Psi}}_u^{(1)}} \ {\hat {\rv{\Psi}}_u^{(2)}} \ \cdots $ $ \ {\hat {\rv{\Psi}}_u^{(n_{\rm{r}})}}  \big{]}^\dag$. The \ac{mse} of the combined normalized squared \ac{af} in \eqref{eq:weight} is expressed as
\begin{align}\label{MMSE}
\mathbb{E}\Big{\{}\big{(}{\widetilde{{\rv{\Psi}}}}_u- {{\Psi}}_u\big{)}^2\Big{\}}=
\V{\Lambda}_{u}^\dag \M{C}_u \V{\Lambda}_{u}+\Big{(}\V{\Lambda}_{u}^\dag \V{\mu}_u-{{\Psi}}_u  \Big{)}^2,
\end{align}
where $\M{C}_u \triangleq  \mathbb{E}\Big{\{} \big{(} \hat {\RV{\Psi}}_u-\V{\mu}_u \big{)} \big{(} \hat {\RV{\Psi}}_u-\V{\mu}_u \big{)}^\dag \Big{\}}$ and
$\V{\mu}_u \triangleq \mathbb{E} \big{\{}\hat {\RV{\Psi}}_u \big{\}}$.}

{By employing the method of Lagrange multipliers, the optimal weighting vector $\V{\Lambda}_{u}^{\rm{op}}$ in \eqref{MMSE} in terms of minimum \ac{mse} is obtained as
\begin{align} \label{eq:optimalw}
\V{\Lambda}_{u}^{\rm{op}}=\big{(}\V{1}^\dag \V{y}_u\big{)}^{-1}\V{y}_u,
\end{align}
where $\V{y}_u \triangleq  \big{(}\M{C}_u+(\V{\mu}_u -{{\Psi}}_u \V{1})(\V{\mu}_u -{{\Psi}}_u \V{1})^\dag\big{)}^{-1} \V{1}$ and $\V{1} \triangleq [1 \ 1\ \cdots \ 1]^\dag$ is an
$n_{\rm{r}}$-dimensional vector of ones.}

{As seen, the optimal weighting vector, $\V{\Lambda}_{u}^{\rm{op}}$, in \eqref{eq:optimalw} depends on the true value of MDS, i.e., $f_{\rm{D}}$, through the true normalized squared AF, ${{\Psi}}_u$, in $\V{y}_u$. To obtain the optimal weighting vector, the mean vector $\V{\mu}_u$ and covariance matrix $\M{C}_u$ are required to be estimated from the received symbols.}
{One approach is bootstrapping \cite{zoubir1998bootstrap,zoubir2007bootstrap,zoubir2000bootstrap}.
The bootstrap method
suggests to re-sample the empirical joint \ac{cdf} of $\hat {\RV{\Psi}}_u$ to estimate $\V{\mu}_u$ and $\M{C}_u$ as summarized in Algorithm \ref{euclidendwhilemmmmm}.\footnote{{Since ${\hat {\rv{\Psi}}_u^{(n)}}$, $n=1,...,n_{\rm{r}}$ are uncorrelated random variables, $\M{C}_u$ is a diagonal matrix. Thus, only the diagonal elements of $\hat{\M{C}}_u$ are employed to obtain the optimal weighting vector.}}}

   \begin{algorithm}[!h]\label{bootstrap}
    \caption{: {Bootstrap Algorithm for Optimal Combining} }
    \begin{algorithmic}[1]
\State {Set $N_{\rm{B}}$}
\For{{$t=1,2,\cdots,N_{\rm{B}}$}}
        \State {{Draw a random sample of size $N$, with replacement, from
        $\Set{X} \triangleq \{1,2,\cdots, N\}$} {and name it $\Set{X}^\star$}}
       % $\RS{X}^{(n)} \triangleq \{\rv{r}_1^{(n)}, \rv{r}_2^{(n)}, \cdots, \rv{r}_N^{(n)}\}$ and
%         $\RS{Y}^{(n)} \triangleq \{ (\rv{r}_1^{(n)},\rv{r}_{1+u}^{(n)}),(\rv{r}_2^{(n)},\rv{r}_{2+u}^{(n)}) \cdots, $ $, (\rv{r}_{N-u}^{(n)},\rv{r}_{N}^{(n)})\}$}
%\State Calculation of the bootstrap estimate: Evaluate the bootstrap estimate of $\hat {\RV{\Psi}}_u^\star$
%as
\For{{$n=1,2,\cdots,n_{\rm{r}}$}} \nonumber
{
\begin{align}
{\hat {\rv{\Psi}}_u^{(n)\star}}[t]=\frac{ \frac{1}{N-u}\sum\limits_{k \in \Set{X}^\star } {\big{|}\rv{r}_k^{(n)}{\big{|}^2}\big{|}\rv{r}_{k +u}^{(n)}{\big{|}^2}}-\Big{(}\frac{1}{N}\sum\limits_{k \in \Set{X}^\star } {|\rv{r}_k^{(n)}{\big{|}^2}}\Big{)}^2}{\frac{1}{N}\sum\limits_{k \in \Set{X}^\star} {|\rv{r}_k^{(n)}{\big{|}^4}}-2\Big{(}\frac{1}{N}\sum\limits_{k \in \Set{X}^\star } {|\rv{r}_k^{(n)}{\big{|}^2}}\Big{)}^2}
\end{align}}
\vspace{-2em}
\EndFor\label{euclidendwhilev}
\State {$\hat {\RV{\Psi}}_u^\star[t] \triangleq 2({\Omega_{\rm{s}}} - 1) \big{[}{\hat {\rv{\Psi}}_u^{(1)\star }} \ {\hat {\rv{\Psi}}_u^{(2)\star }} \ \cdots \ {\hat {\rv{\Psi}}_u^{(n_{\rm{r}})\star }}  \big{]}^\dag$}
\EndFor\label{euclidendwhilev}
\State {${\RM{\Gamma}}_u=\Big{[}\hat {\RV{\Psi}}_u^\star[1] \ \hat {\RV{\Psi}}_u^\star[2] \ \cdots \ \hat {\RV{\Psi}}_u^\star[N_{\rm{B}}]\Big{]}$}
\State {${\hat{\V{\mu}}_u} = \frac{1}{N_{\rm{B}}} \sum_{t=1}^{N_{\rm{B}}} \hat {\RV{\Psi}}_u^\star[t]$}
\State {$\hat{\M{C}}_u=\frac{1}{N_{\rm{B}}-1}({\RM{\Gamma}}_u-\hat{\V{\mu}}_u \V{1}^\dag)({\RM{\Gamma}}_u-\hat{\V{\mu}}_u \V{1}^\dag)^\dag$}
    \end{algorithmic}\label{euclidendwhilemmmmm}
  \end{algorithm}

{As seen in Algorithm \ref{euclidendwhilemmmmm}, the optimal weighting vector for each delay
lag $u$ is derived at the expense of higher computational complexity. In order to avoid this computational complexity, the suboptimal equal weight combining method can be employed as }
\begin{align}\label{eq:weightrrr}
{\widetilde{{\rv{\Psi}}}}_u = \frac{1}{n_{\rm{r}}}\sum\limits_{n = 1}^{{n_{\rm{r}}}} {\hat {\rv{\Psi}}_u^{(n)}}.
\end{align}

\begin{figure}[!t]
\centering
\includegraphics[width=3.7in]{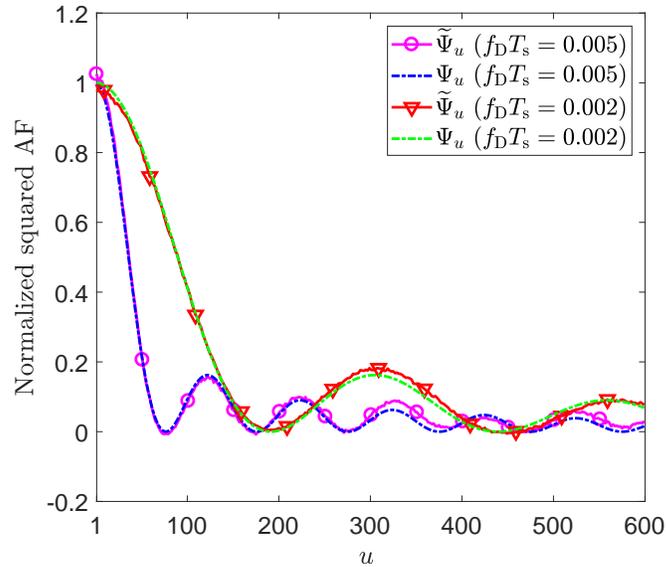}
\vspace{-1em}
\caption{Illustration of the non-linear LS regression for the uniformly sampled normalized squared AF for $f_{\rm{D}}T_{\rm{s}}=0.02$  and $f_{\rm{D}}T_{\rm{s}}=0.005$, with $n_{\rm{t}}=1$, $n_{\rm{r}}=2$, $L=1$, $u_{\rm{s}}=2$, and at $\gamma=10$ dB.}\label{fig:vvb}
\end{figure}

Fig. \ref{fig:vvb} shows how ${\widetilde{{\rv{\Psi}}}}_u$ fits ${{\Psi }_{u}}$ through the equal weight combining in \eqref{eq:weightrrr}  for $f_{\rm{d}}T_{\rm{s}}=0.02$  and $f_{\rm{d}}T_{\rm{s}}=0.005$  with $n_{\rm{t}}=1$, $n_{\rm{r}}=2$, $L=1$, $u_{\rm{s}}=2$, and at $\gamma=10$ dB.

Finally, similar to the MISO scenario, the problem of \ac{mds} estimation for multiple receive antennas is formulated as non-linear regression problem in \eqref{eq:TT23} for ${\widetilde{{\rv{\Psi}}}}_u$.
A formal description
of the proposed  \ac{nda}-\ac{mbe} for \ac{mds} in \ac{mimo} frquency-selective channel is presented in Algorithm \ref{al:3}.
  \begin{algorithm}[]
    \caption{: \ac{nda}-\ac{mbe} for \ac{mds} in  MIMO systems}\label{euclid}
    \begin{algorithmic}[1]
\State {Set $f_{\rm{l}}$, $f_{\rm{h}}$, $\Delta$, and $\delta$ }
\State {Acquire the measurements $\{ {{\rv{r}_k^{(n)}}} \}_{k = 1}^N$,\,\,\ $\forall n \in \big{\{}1,\hdots,n_{\rm{r}}\big{\}}$}
\State {Estimate the statistics $\hat {\rv{\mu}}_2^{(n)}$, $\hat {\rv{\mu}}_4^{(n)}$, and $\hat {\rv{\kappa}}_u^{(n)}$, by employing \eqref{eq:TT14} for $\{ {{\rv{r}_k^{(n)}}} \}_{k = 1}^N$,\,\,\ $\forall n \in \big{\{}1,\hdots,n_{\rm{r}}\big{\}}$}
\State {Compute ${{\hat {\rv{\Psi}} }_u^{(n)}}$, $\forall u \in \big{\{}U_{\rm{min}},\hdots,U_{\rm{max}}\big{\}}$, $\forall n \in \big{\{}1,\hdots,n_{\rm{r}}\big{\}}$,
 by using \eqref{eq:TT19}}
\State {Compute ${\widetilde{{\rv{\Psi}}}}_u$, $\forall u \in \big{\{}U_{\rm{min}},\hdots,U_{\rm{max}}\big{\}}$, by using \eqref{eq:weightrrr}}
\State {Obtain $\hat {\rv{F}}_{\rm{D}}^{(\rm{r})}$ by solving the minimization problem in \eqref{eq:TT20} for ${\widetilde{{\rv{\Psi}}}}_u$ through the grid search method with step size $\Delta$}
\State {Obtain $\hat {\rv{F}}_{\rm{D}}^{(\rm{s})}$ by solving the minimization problem in \eqref{eq:TT20}
for ${\widetilde{{\rv{\Psi}}}}_u$
via the grid search method over $\big{[}\hat {\rv{F}}_{\rm{D}}^{(\rm{r})}-\Delta,\hat {\rv{F}}_{\rm{D}}^{(\rm{r})}+\Delta\big{]}$ with step size $\delta$}
\State {$\hat{\rv{F}}_{\rm{D}}=\hat{\rv{F}}_{\rm{D}}^{(\rm{s})}$}
    \end{algorithmic}\label{al:3}
  \end{algorithm}

\subsection{Semi-blind \ac{nda}-\ac{mbe}}
The proposed \ac{nda}-\ac{mbe} for  MISO and \ac{mimo} systems do not require knowledge of
the parameter vector $\V{\varphi}= [\V{\beta}^\dag \ $ $ \V{{\xi}}^\dag \V{{\vartheta}}^\dag \ f_{\rm{D}}]^\dag$.
In other words, the proposed \ac{nda}-\ac{mbe} in section \ref{section:mba} and \ref{section:mbb} are blind.
For the scenarios in which the variance of the additive noise can be accurately estimated at the receive antennas, i.e., $\V{{\xi}}$ is known,  a semi-blind \ac{nda}-\ac{mbe} for the case of \ac{siso} transmission and flat-fading channel, i.e., $n_{\rm{t}}=1$ and $L=1$, can be proposed.
In this case, for the $n$th receive antennas, one can easily obtain\footnote{The index of transmit antenna, i.e., $m=1$ and the index of channel tap, i.e., $l=1$ is dropped.}
\begin{align}\label{eq:rmard1}
{\mu _2^{(n)}} = \sigma_{{\rm{h}}_n}^2\sigma_{\rm{s}}^2 + \sigma_{{\rm{w}}_n}^2
\end{align}
and
\begin{align}\label{eq:rmard2}
\eta^{(n)}= (\sigma_{{\rm{h}}_n}^{4}\sigma_{\rm{s}}^{4})^{-1}= \frac{1}{\Big{(}{\mu _2^{(n)}} - \sigma_{{\rm{w}}_n}^2\Big{)}^{2}}.
\end{align}
By using \eqref{eq:TT11y}, \eqref{eq:rmard1} and \eqref{eq:rmard2}, and by replacing the statistical moments and the noise variance with their corresponding estimates, one obtains
\begin{equation}\label{eq:n27}
{{\hat {\rv{\Psi}} }_u^{(n)}} = \frac{{{{\hat {\rv{\kappa}} }_u^{(n)}} - \Big{(}\hat {\rv{\mu}}_2^{(n)}\Big{)}^2}}{{\Big{(}{{\hat {\rv{\mu}} }_2^{(n)}} - \hat \sigma _{\rm{w}}^2\Big{)}^2}},
\end{equation}
where $\hat \sigma_{{\rm{w}}_n}^2$ is the estimate of the noise variance, and $\hat {\rv{\kappa}}_u^{(n)}$ and $\hat {\rv{\mu}}_2^{(n)}$ are given in \eqref{eq:TT14}. Clearly, similar to the SISO transmission, the optimal and suboptimal combining methods for the multiple receive antennas can be employed, as well.

\begin{table*}[!t]
\centering
\caption{Number of real additions, real multiplications, and complexity order of the proposed \ac{nda}-\ac{mbe}.}
\label{my-label}
{\small
\begin{tabular}{lllll}
                        &                        &                        &                        &  \\ \cline{1-4}
\multicolumn{1}{|l|}{Algorithm} & \multicolumn{1}{l|}{\,\,\,\,\,\,\,\,\,\,\,\,\,\,\,\,\,\,\,\,\,\,\,\,\,\,\,\,\,\,\,\,\,\,\,\,\,\,\,\,\,\,\,\,\,\,\,\,\,\,\,\, Real additions} & \multicolumn{1}{l|}{\,\,\,\,\,\,\,\,\,\,\,\,\,\,
\,\,\,\,\,\,\,\,\,\,\,\,\,\,\,\,\,\,\,\,\,\,\,\,\,\,\,\,\,\,\,\,
Real multiplications} & \multicolumn{1}{l|}{Order} &  \\ \cline{1-4}
\multicolumn{1}{|l|}{\,\,\,\,\ MISO} & \multicolumn{1}{l|}{\,\,\,\,\,\,\,\,\,\ $
\Big{(}N+2N_{\rm{g}}-\frac{(U_{\rm{max}}+U_{\rm{min}})}{2}\Big{)}N_{\rm{la}}+3N-N_{\rm{g}}-1$}& \multicolumn{1}{l|}{\,\,\,\,\,\,\,\,\,\,\,\ $\Big{(}N+N_{\rm{g}}-\frac{(U_{\rm{max}}+U_{\rm{min}})}{2}+2\Big{)}N_{\rm{la}}+3N+4$} & \multicolumn{1}{l|}{$\mathcal{O}(N)$} &  \\ \cline{1-4}
\multicolumn{1}{|l|}{\,\,\,\ MIMO} & \multicolumn{1}{l|}{$n_{\rm{r}}\Big{(}\big{(}N-\frac{(U_{\rm{max}}+U_{\rm{min}})}{2}\big{)}N_{\rm{la}}+3N-1\Big{)}+(2N_{\rm{la}}-1)N_{\rm{g}}$} & \multicolumn{1}{l|}{$n_{\rm{r}}\Big{(}\big{(}N-\frac{(U_{\rm{max}}+U_{\rm{min}})}{2}+2\big{)}N_{\rm{la}}+3N+4\Big{)}+N_{\rm{g}}N_{\rm{la}}+1$} & \multicolumn{1}{l|}{$\mathcal{O}(N)$} &  \\ \cline{1-4}
\end{tabular}\label{table}
}
\end{table*}

\begin{figure}
\centering
\includegraphics[width=3.7in]{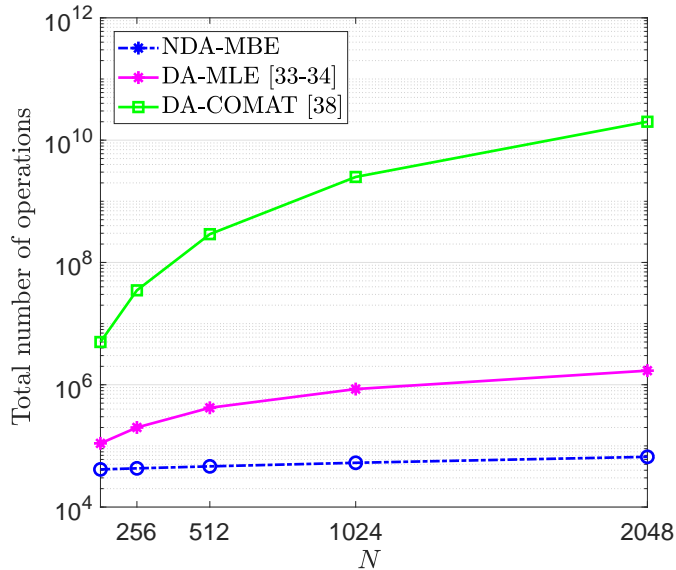}
\caption{Computational complexity comparison of the proposed \ac{nda}-\ac{mbe}, the low-complexity \ac{da}-\ac{mle} in \cite{bellili2013low,bellili2017low}, and the \ac{da}-COMAT estimator in \cite{R225}.}\label{fig_erp}
\end{figure}
\begin{figure}[!b]
\centering
 \vspace{-1em}
 \subfigure[$f_{\rm{D}}=1000$ Hz]{%
    \includegraphics[width=.45\textwidth]{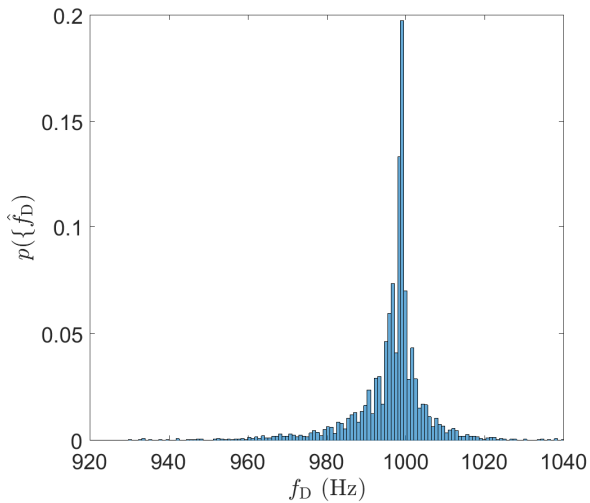} \label{fig:1}
  }
  \\
  \subfigure[$f_{\rm{D}}=100$ Hz]{%
   \includegraphics[width=.45\textwidth]{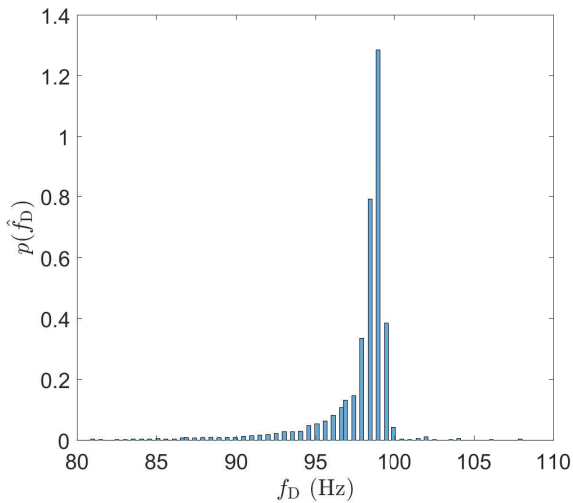} \vspace{-1em} \label{Figures/fig:2}
  }
  \caption{Distribution of the estimated $\hat f_{\rm{D}}$ for the proposed \ac{nda}-\ac{mbe} for $n_{\rm{t}}=2$, $n_{\rm{r}}=2$, and at $\gamma=10$ dB. }\label{fig_erpty}
\end{figure}

{\section{Complexity Analysis}\label{section:complexity}
By employing the two steps grid search method to solve the optimization problem in \eqref{eq:TT20}, the number of real additions and multiplications employed in the proposed \ac{nda}-\ac{mbe} is
shown in Table \ref{table}, where $N_{\rm{la}}$ is the number of delay lag, $N_{\rm{g}}\triangleq (N_{\rm{g}_1}+N_{\rm{g}_2})$, and $N_{\rm{g}_1}$ and $N_{\rm{g}_2}$ are the
number of grid points used for the rough and fine estimation, respectively. As seen, the proposed \ac{nda}-\ac{mbe} exhibits a complexity order of $\mathcal{O}(N)$. It should be mentioned that the complexity order of the derived \ac{da}-\ac{mle} and \ac{nda}-\ac{mle} are $\mathcal{O}(N^3)$ and $\mathcal{O}(|M|^{N'n_{\rm{t}}})$, respectively.}

{Fig. \ref{fig_erp} compares the total number of operations used by the proposed \ac{nda}-\ac{mbe} with the low-complexity \ac{da}-\ac{mle} in \cite{bellili2013low,bellili2017low} and the \ac{da}-COMAT estimator in \cite{R225}. As seen, the proposed
\ac{nda}-\ac{mbe} exhibits significantly lower computational complexity compared to the \ac{da}-COMAT \cite{R225} and the low-complexity \ac{da}-\ac{mle} in \cite{bellili2013low,bellili2017low}.
This substantial reduced-complexity enables the proposed \ac{mbe} to exhibit good performance in the \ac{nda} scenarios,  where the observation window can be selected large enough.}
%{Fig. \ref{fig_erp} compares the total number of operations used by the proposed \ac{nda}-\ac{mbe} with the low-complexity \ac{da}-\ac{mle} in \cite{bellili2013low,bellili2017low} and the \ac{da}-COMAT estimator in \cite{R225}. As seen, the \ac{nda}-\ac{mbe} exhibits significantly lower computational complexity compared to the \ac{da}-COMAT and estimator. Furthermore, the computational complexity of the proposed \ac{mbe} is close to the \ac{mds} estimator in \cite{bellili2013low,bellili2017low}.}

\section{Simulation Results}\label{section:simulation}
In this section, we examine the performance of the proposed \ac{nda}-\ac{mbe}, as well as the derived
 \ac{da}-\ac{mle} and  \ac{da}-\ac{crlb} for \ac{mds} in \ac{mimo} frequency-selective fading channel  through several simulation
experiments.

\subsection{Simulation Setup}
We consider a \ac{mimo} system employing spatial multiplexing, with carrier frequency  $f_{\rm{c}}=2.4$ GHz. Unless otherwise mentioned, $n_{\rm{t}}=2$, $n_{\rm{r}}=2$, $T_{\rm{s}}=10 \,\mu s$, $N=10^{5}$, and the modulation is 64-QAM.
The delay profile of the Rayleigh fading channel is $\sigma_{{{\rm_h}_{(mn),l}^{}}}^2=\beta\exp \left( {{{ - {\l _{{\rm{rms}}}}l} \mathord{\left/
 {\vphantom {{ - {\tau _{\rm{rms}}}l} L}} \right.
 \kern-\nulldelimiterspace} L}} \right)$,
where $\beta$ is a normalization factor, i.e., $\beta\sum\nolimits_l {({{ - {l_{{\rm{rms}}}}l} \mathord{\left/
 {\vphantom {{ - {l_{{\rm{rms}}}}l} {L)}}} \right.
 \kern-\nulldelimiterspace} {L)}}}  = 1$, with $L=5$
and ${{\l _{\rm{{rms}}}}}=L/4$ as the maximum and RMS delay spread of the channel, respectively. The parameters for the downsampled LS curve-fitting optimization are ${U_{\min }}=L$, ${U_{\max }}=\lfloor \frac{N}{10}\rfloor$, and $u_{\rm{s}}=10$. The additive white noise was modeled as a complex-valued Gaussian random variable
 with zero-mean and variance $\sigma_{\rm{w}}^2$ for each receive antennas.
 Without loss of generality, it was assumed that $\sigma_{{\rm{s}}_{{m}}}^2=1/(n_{\rm{t}}n_{\rm{r}})$, $m=1,2,..,n_{\rm{t}}$, and thus, the
average SNR was defined as $\gamma= 10\log ({1 \mathord{\left/
 {\vphantom {1 {{n_{\rm{r}}}\sigma_{\rm{w}}^2}}} \right.
 \kern-\nulldelimiterspace} {{n_{\rm{r}}}\sigma_{\rm{w}}^2}})$.
 {
Unless otherwise mentioned,
the performance of the \ac{mds} estimators was presented in terms of normalized \ac{nrmse}, i.e., $\mathbb{E}{\{{({{\hat f}_{\rm{D}}T_{\rm{s}}} - {f_{\rm{D}}}T_{\rm{s}})^2}\}^{1/2}}/f_{\rm{D}}T_{\rm{s}}$, obtained from 1000 Monte Carlo trials for each
${f_{\rm{D}}T_{\rm{s}}} \in [10^{-3}, 18 \times 10^{-3}]$, with the search step size $ \Delta=10$ Hz and $\delta=0.5$ Hz, respectively.}

\subsection{Simulation Results}
\begin{figure}
\centering
\includegraphics[width=3.7in]{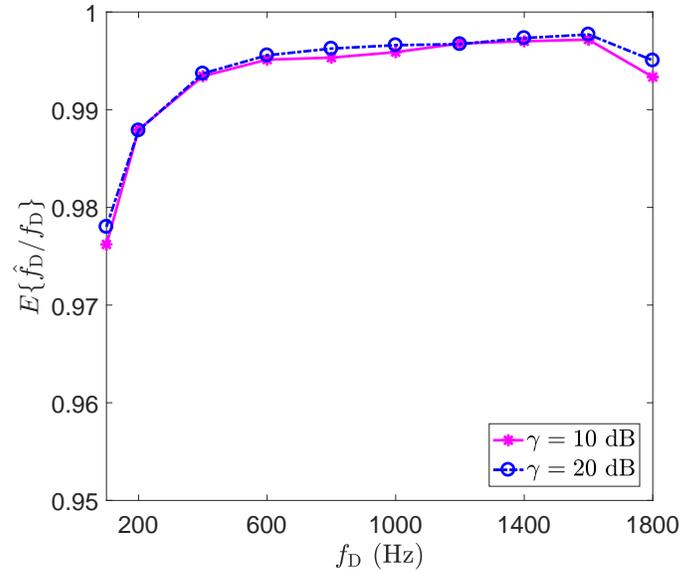}
\caption{The mean value of the estimated \ac{mds} by \ac{nda}-\ac{mbe} for various SNR values for $n_{\rm{t}}=2$ and $n_{\rm{r}}=2$.}\label{fig:mio}
\end{figure}

{Fig. \ref{fig_erpty} shows the distributions of the estimated $\hat f_{\rm{D}}$ by the proposed \ac{nda}-\ac{mbe} for different \ac{mds}s, $ f_{\rm{D}}= 1000$ Hz and $f_{\rm{D}}= 100$ Hz, with $n_{\rm{t}}=2$, $n_{\rm{r}}=2$, and at $\gamma=10$ dB. As seen, the distributions are not symmetric around their mean values; hence, this leads to bias in \ac{mds} estimation.} Furthermore,
Fig. \ref{fig:mio} illustrates $\mathbb{E}\{\hat{f_{\rm{D}}}/f_{\rm{D}}\}$ versus $f_{\rm{D}}$ for $\gamma=10$ dB and $\gamma=20$ dB. As seen, the proposed \ac{nda}-\ac{mbe} is nearly unbiased, i.e., $\mathbb{E}\{\hat{f_{\rm{D}}}\}\approx f_{\rm{D}}$ over a wide range of \ac{mds}. This can be explained, as while the distribution of the estimated $f_{\rm{D}}$ is not symmetric, the estimated values are accumulated around their mean value. It should be mentioned that by increasing the length of the observation window, $N$, the bias of the proposed estimator approaches zero.
\begin{figure}
\centering
\includegraphics[width=3.7in]{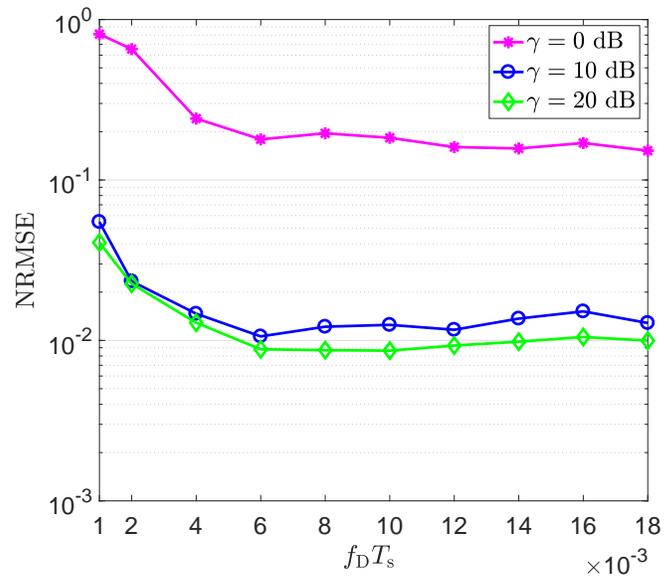}
\caption{The NRMSE of the proposed \ac{nda}-\ac{mbe} versus $f_{\rm{D}}T_{\rm{s}}$ for different SNR values.}\label{fig:miotyt}
\end{figure}
\begin{figure}
\centering
\includegraphics[width=3.7in]{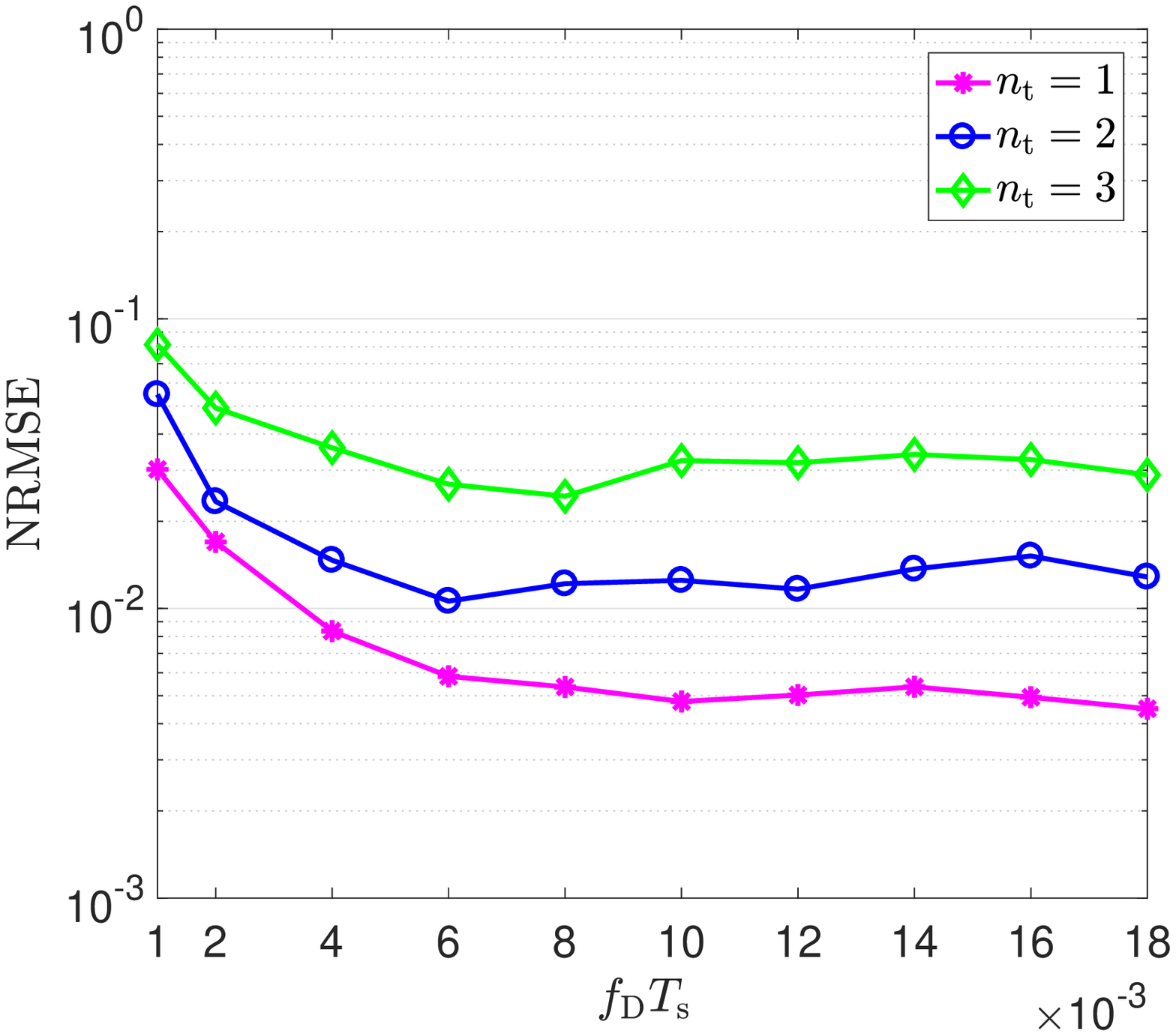}
\caption{The effect of $n_{\rm{t}}$ on the performance
of the proposed \ac{nda}-\ac{mbe}  for $n_{\rm{r}}=2$ and at $\gamma=10$ dB.}\label{fig:miklko}
\end{figure}
\begin{figure}[!t]
\centering
\includegraphics[width=3.7in]{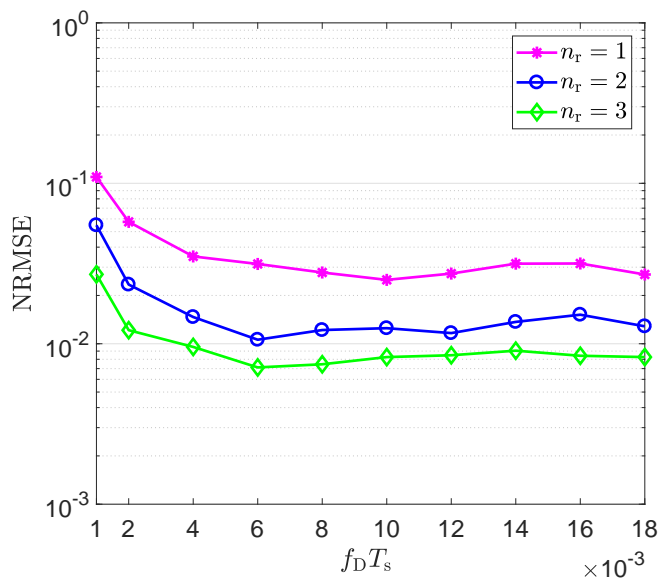}
\caption{The effect of $n_{\rm{r}}$ on the performance of the proposed \ac{nda}-\ac{mbe} for $n_{\rm{t}}=2$ and at $\gamma=10$ dB.}\label{fig:llpliomiottte}
\end{figure}

In Fig. \ref{fig:miotyt}, the \ac{nrmse} of the \ac{nda}-\ac{mbe} versus $f_{\rm{d}}T_s$ is illustrated for $\gamma=0$ dB, $\gamma=10$ dB, and $\gamma=20$ dB. As seen, the proposed estimator exhibits a good performance over a wide range of Doppler rates, $f_{\rm{D}}T_{\rm{s}}$. As observed, the \ac{nrmse} decreases as $f_{\rm{D}}T_{\rm{s}}$  increases.
This performance improvement can be explained, as for lower Doppler rates, a larger observation window is required to capture the variation of the fading channel.
Also, as expected, the \ac{nrmse} decreases as $\gamma$ increases. This can be easily explained, as
an increase in $\gamma$ leads to more accurate estimates of the statistics in \eqref{eq:TT19}.

Fig. \ref{fig:miklko} presents the \ac{nrmse} of the proposed \ac{nda}-\ac{mbe} versus $f_{\rm{D}}T_s$ for different numbers of transmit antennas, $n_{\rm{t}}$, for $n_{\rm{r}}=2$ and at $\gamma=10$ dB. As expected, the \ac{nrmse} increases as the number of transmit antennas increases. This increase can be explained, as the variance of the statistics employed in \eqref{eq:TT19} increases with the number of transmit antennas, thus, leading to higher estimation error in the LS curve-fitting.

In Fig. \ref{fig:llpliomiottte}, the \ac{nrmse} of the proposed \ac{nda}-\ac{mbe} is shown versus $f_{\rm{D}}T_{\rm{s}}$ for different numbers of receive antennas, $n_{\rm{r}}$, for $n_{\rm{t}}=2$, and at $\gamma=10$ dB. It can be seen that an increment in $n_{\rm{r}}$ leads to a reduced NRMSE. This decrease can be easily explained, as averaging at the receive-side yields more accurate estimation of ${ {{\Psi}} _u}$, thus, leading to a more accurate result
in the LS curve-fitting.

\begin{figure}[!t]
\centering
\includegraphics[width=3.7in]{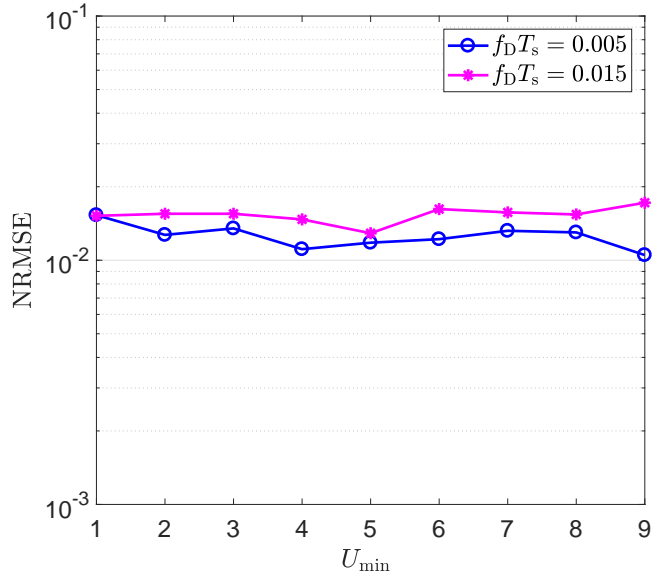}
\caption{The effect of the parameter ${U_{\min }}$ on the performance of the proposed \ac{nda}-\ac{mbe} for different values of $f_{\rm{d}}T_{\rm{s}}$ and at $\gamma=10$ dB.} \label{fig:llpliomio}
\end{figure}

In Fig. \ref{fig:llpliomio}, the effect of the parameter ${U_{\min }}$ on the performance of the proposed \ac{nda}-\ac{mbe} is illustrated for $L=5$ and $u_{\rm{s}}=10$. As observed, the proposed estimator exhibits a low sensitivity to the value of ${U_{\min }}$. This can be explained, as a large number of lags, ${U_{\min }} \le u \le {U_{\max }}$, are employed for fitting ${{\widetilde {\rv{\Psi}} }_u}$ to $J_0^2(2\pi {f_{\rm{D}}}{T_{\rm{s}}}u)$ in the LS estimation; thus, the estimator is nearly robust to a few missing delay lags, $L \le u < {U_{\min }}$, or nuisance delay lags, $1 \le u < L$. As such, basically the estimator does not require an accurate estimate of $L$.

\begin{figure}[!t]
\centering
\includegraphics[width=3.7in]{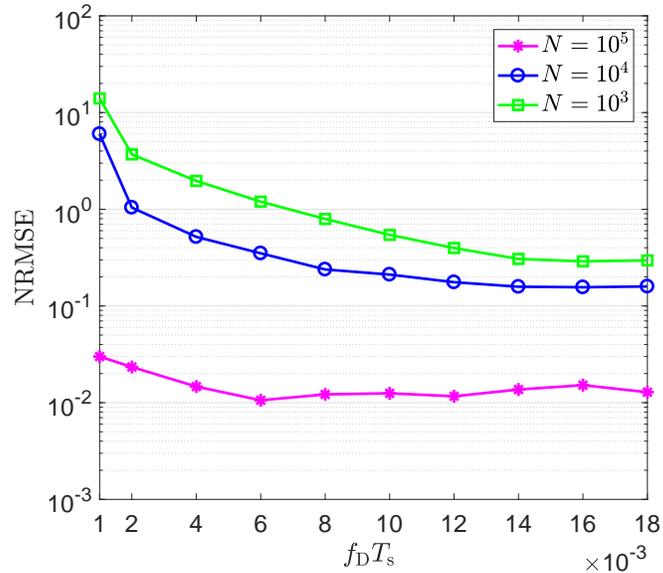}
\caption{The effect of
 the observation window size, $N$,  on the performance of the proposed \ac{nda}-\ac{mbe} for $n_{\rm{t}}=2$ and  $n_{\rm{r}}=2$ in frequency-selective channel, and at $\gamma=10$ dB.}\label{fig:56}
\end{figure}

{Fig. \ref{fig:56} shows the effect of
 the observation window size, $N$, on the performance of the proposed \ac{nda}-\ac{mbe}. As expected, the performance of the proposed estimator improves as the length of the observation window increases. This performance improvement can be explained, as the variance of the estimated statistics employed in \eqref{eq:TT19} decreases when $N$ increases.}

\begin{figure}[!t]
\centering
\includegraphics[width=3.7in]{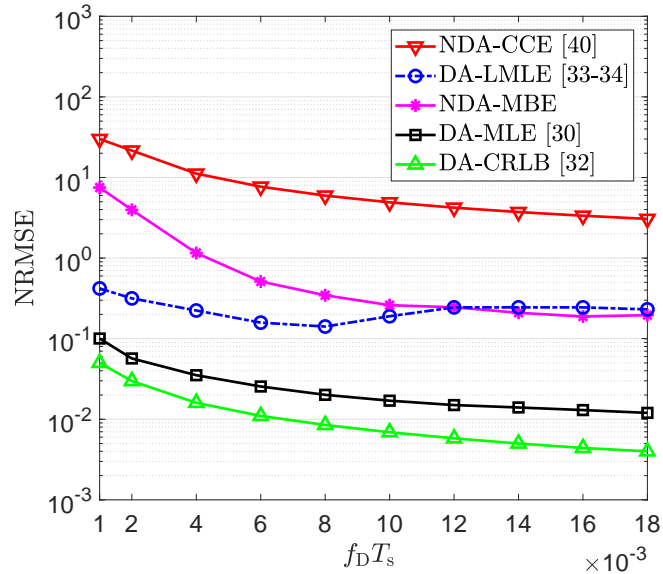}
\caption{Performance comparison of the proposed NDA-MBE, DA-CRLB \cite{R18}, DA-MLE \cite{R22}, the low-complexity DA-MLE in \cite{bellili2013low,bellili2017low}, and the NDA-CCE in \cite{R23} in \ac{siso} frequency-flat fading channel for $N=10^3$ at $\gamma=10$ dB.}\label{fig:windowst}
%\cite{R15}
\end{figure}

{In Fig. \ref{fig:windowst}, the \ac{nrmse} is plotted versus $f_{\rm{D}}T_{\rm{s}}$ for the proposed \ac{nda}-\ac{mbe}, the low-complexity \ac{da}-\ac{mle} (\ac{da}-LMLE) in \cite{bellili2017low,bellili2013low}, the \ac{nda}-\ac{cce} in \cite{R23}, the \ac{da}-\ac{mle} in \cite{R22}, and the DA-CRLB in \cite{R18} for \ac{mds} estimation in \ac{siso} frequency-flat fading channel for $N=1000$ and
 at $\gamma=10$ dB. As seen, the proposed \ac{nda}-\ac{mbe} outperforms the \ac{nda}-\ac{cce}, and provides a similar performance as the \ac{da}-LMLE  for
$f_{\rm{D}}T_{\rm{s}}\geq 0.012$. The performance degradation of the \ac{da}-LMLE at high values of $f_{\rm{D}}T_{\rm{s}}$ is related to
the second-order Taylor expansion employed to approximate the covariance matrix; this is less accurate at higher \ac{mds}s.
}
\begin{figure}
\centering
\includegraphics[width=3.7in]{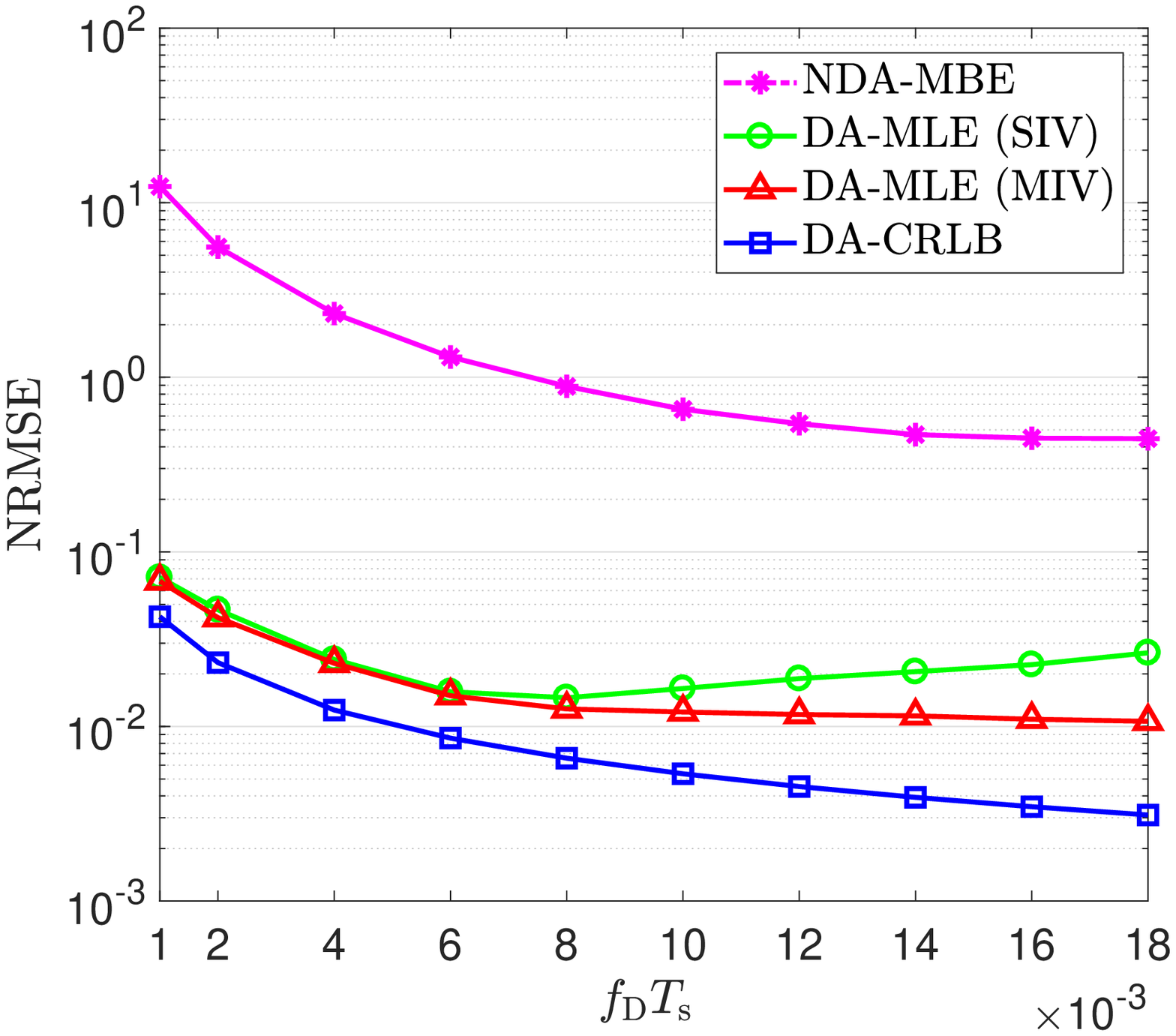}
\caption{Performance comparison of the proposed NDA-MBE, the derived DA-CRLB, and the derived DA-MLE (SIV and MIV) in \ac{mimo} frequency-selective Rayleigh fading channel for $n_{\rm{t}}=2$, $n_{\rm{r}}=2$, $L=5$, $N=10^3$, and  at $\gamma=10$ dB.}\label{fig:windowsterc908}
\end{figure}

{Fig. \ref{fig:windowsterc908} illustrates the NRMSE versus $f_{\rm{D}}T_{\rm{s}}$ for the proposed \ac{nda}-\ac{mbe}, the derived \ac{da}-\ac{mle}, and the derived \ac{da}-\ac{crlb} in \ac{mimo} frequency-selective fading channel for $N=1000$ and at $\gamma=10$ dB.
{In order to show the convergence problem in the derived \ac{da}-\ac{mle} caused by the Fisher-scoring numerical method employed to solve the \ac{ml} \cite{kay1993fundamentals},
the performance of the derived \ac{da}-\ac{mle} for the cases of single initial value (SIV) and multiple initial values (MIV) is plotted, respectively.\footnote{{With the MIV method, several initial values are considered and at convergence the
one that yields the maximum is chosen.}}
As seen, by choosing MIV, the convergence problem of the Fisher-scoring method employed in the derived \ac{da}-\ac{mle} is solved.
Moreover, as observed, the performance of the derived \ac{da}-\ac{mle} with MIV
is close to the \ac{da}-\ac{crlb}.} This high performance is obtained at the expense of significant computational
complexity in the order of $\mathcal{O}(N^3)$. On the other hand, the proposed \ac{nda}-\ac{mbe}
cannot reach the \ac{da}-\ac{crlb}. This behaviour can be explained, as the \ac{nda}-\ac{mbe} requires a larger number of
observation symbols to accurately estimate the second- and fourth-order statistics in time-varying channel.
However, the substantial reduced-complexity enables the proposed \ac{mbe} to exhibit significantly low \ac{nrmse} in the \ac{nda} scenarios,  where the observation window can be selected large enough.}
%\textcolor{blue}{Moreover, as seen, by choosing MIV, the convergence problem of the Fisher-scoring method employed in the derived \ac{da}-\ac{mle} is solved.}

\begin{figure}
\centering
\includegraphics[width=3.7in]{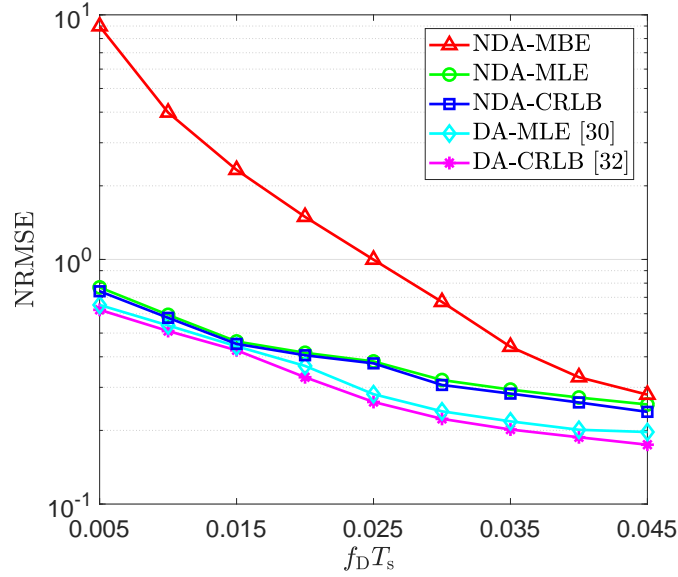}
\caption{Performance comparison of the proposed semi-blind \ac{nda}-\ac{mbe} in \eqref{eq:n27}, the derived \ac{nda}-\ac{mle}, the derived \ac{nda}-\ac{crlb}, the \ac{da}-\ac{mle} in \cite{R22}, and the \ac{da}-\ac{crlb} in \cite{R18} in \ac{siso} frequency-flat-fading channel for $N=10$ and  at $\gamma=20$  dB SNR.}\label{fig:windowsterc908o0}
\end{figure}

{In Fig. \ref{fig:windowsterc908o0}, the \ac{nrmse} is shown versus $f_{\rm{D}}T_{\rm{s}}$ for the proposed semi-blind \ac{nda}-\ac{mbe} in \eqref{eq:n27}, the derived
\ac{nda}-\ac{mle}, and the \ac{nda}-\ac{crlb} in \ac{siso} flat-fading channel for BPSK signal, $N=10$, { $f_{\rm{D}}T_{\rm{s}} \in [5 \times 10^{-3},45 \times 10^{-3}]$}, and at $\gamma=20$ dB.\footnote{The complexity order of the derived \ac{nda}-\ac{mle} and \ac{nda}-\ac{crlb} are in the order of $\mathcal{O}(|M|^{N'n_{\rm{t}}})$; for large values of $N'$ ($N'=N+L-1$), the corresponding curves are not obtainable even for $|M|=2$ or $n_{\rm{t}}=1$. Hence, $N=10$ and \ac{siso} flat-fading channel are considered.}
As expected, the \ac{nda}-\ac{mbe} does not exhibit good performance for a short observation window size because
the second- and fourth-order statistics employed in \eqref{eq:n27} are not accurately estimated.  On the other hand,
the derived \ac{nda}-\ac{mle} exhibits low \ac{nrmse} even for a short observation window. Moreover, there is no significant performance gap between the derived \ac{nda}-\ac{mle}
and \ac{nda}-\ac{crlb},  as well as the \ac{da}-\ac{mle} in  \cite{R22} and the \ac{da}-\ac{crlb} in \cite{R18}.}

\begin{figure*}[!t]
\normalsize
\begin{align}\label{uuu}
\big{|}\rv{r}_k^{(n)}{\big{|}^2} &=\sum\limits_{m = 1}^{{n_{\rm{t}}}} {\sum\limits_{l = 1}^L {\big{|}\rv{h}_{k,l}^{(mn)}{\big{|}^2}} }\big{|}\rv{s}_{k - l}^{(m)}{\big{|}^2} + \sum\limits_{{m_1} = 1}^{{n_{\rm{t}}}} {\sum\limits_{{m_2} \ne {m_1}}^{{n_{\rm{t}}}} {\sum\limits_{l = 1}^L {\rv{h}_{k,l}^{({m_1n})}{{\left(\rv{h}_{k,l}^{({m_2n})}\right)^*}}\rv{s}_{k - l}^{({m_1})}{{\left(\rv{s}_{k - l}^{({m_2})}\right)^*}}} } } \\ \nonumber
&\quad + \sum\limits_{m = 1}^{{n_{\rm{t}}}} {\sum\limits_{{l_1} = 1}^L {\sum\limits_{{l_2} \ne {l_1}}^L {\rv{h}_{k,{l_1}}^{(mn)}{{\left(\rv{h}_{k,{l_2}}^{(mn)}\right)^*}}\rv{s}_{k - {l_1}}^{(m)}{{\left(\rv{s}_{k - {l_2}}^{(m)}\right)^*}}} } }
 + \sum\limits_{{m_1} = 1}^{{n_{\rm{t}}}} {\sum\limits_{{m_2} \ne {m_1}}^{{n_{\rm{t}}}} {\sum\limits_{{l_1} = 1}^L {\sum\limits_{{l_2} \ne {l_1}}^L {\rv{h}_{k,{l_1}}^{({m_1n})}{{\left(\rv{h}_{k,{l_2}}^{({m_2n})}\right)^*}}\rv{s}_{k - {l_1}}^{({m_1})}{{\left(\rv{s}_{k - {l_2}}^{({m_2})}\right)^*}} } } } } \\ \nonumber &\quad+{\Big{(}\rv{w}_k^{(n)}\Big{)}^*}\sum\limits_{m = 1}^{{n_{\rm{t}}}} {\sum\limits_{l = 1}^L {\rv{h}_{k,l}^{(mn)}\rv{s}_{k - l}^{(m)}} }
+\Big{(}\rv{w}_k^{(n)}\Big{)}\sum\limits_{m = 1}^{{n_{\rm{t}}}} {\sum\limits_{l = 1}^L {{{\left(\rv{h}_{k,l}^{(mn)}\right)^*}}{{\left(\rv{s}_{k - l}^{(m)}\right)^*}}} }  + \big{|}\rv{w}_k^{(n)}{\big{|}^2}.
\end{align}
\begin{align}\nonumber
\kappa_u^{(n)} &= \sum\limits_{m = 1}^{{n_{\rm{t}}}} {\sum\limits_{l = 1}^L {\,\mathbb{E}\left\{\big{|}\rv{h}_{k,l}^{(mn)}{\big{|}^2}|\rv{h}_{k + u,l}^{(mn)}{\big{|}^2}\right\}} }
\mathbb{E}\left\{\big{|}\rv{s}_{k - l}^{(m)}{\big{|}^2}\big{|}\rv{s}_{k + u - l}^{(m)}{\big{|}^2}\right\} +
\sum\limits_{{m_1} = 1}^{{n_{\rm{t}}}} {\sum\limits_{{m_2} \ne {m_1}}^{n_{\rm{t}}} {\sum\limits_{{l} = 1}^L {\mathbb{E}\left\{\big{|}\rv{h}_{k,{l}}^{({m_1n})}{\big{|}^2}|\rv{h}_{k + u,{l}}^{({m_2n})}{\big{|}^2}\right\}} } } \mathbb{E}\left\{\big{|}\rv{s}_{k - l}^{({m_1})}{\big{|}^2}\big{|}\rv{s}_{k + u - l}^{({m_2})}{\big{|}^2}\right\}\\ \nonumber
& \quad + \sum\limits_{m = 1}^{{n_{\rm{t}}}} {\sum\limits_{{l_1} = 1}^L {\sum\limits_{{l_2} \ne {l_1}}^L {\mathbb{E}\left\{\big{|}\rv{h}_{k,{l_1}}^{(mn)}{\big{|}^2}\big{|}\rv{h}_{k + u,{l_2}}^{(mn)}{\big{|}^2}\right\}} } } \mathbb{E}\left\{\big{|}\rv{s}_{k - {l_1}}^{(m)}{\big{|}^2}\big{|}\rv{s}_{k + u - {l_2}}^{(m)}{\big{|}^2}\right\} \\ \nonumber
& \quad
+\sum\limits_{{m_1} = 1}^{{n_{\rm{t}}}} {\sum\limits_{{m_2} \ne {m_1}}^{{n_{\rm{t}}}} {\sum\limits_{{l_1} = 1}^L {\sum\limits_{{l_2} \ne {l_1}}^L {\mathbb{E}\left\{\big{|}\rv{h}_{k,{l_1}}^{({m_1n})}{\big{|}^2}\big{|}\rv{h}_{k + u,{l_2}}^{({m_2n})}{\big{|}^2}\right\}\mathbb{E}\left\{\big{|}\rv{s}_{k - {l_1}}^{({m_1})}{\big{|}^2}\big{|}\rv{s}_{k + u - {l_2}}^{({m_2})}{\big{|}^2}\right\}} } } }+\mathbb{E}\left\{\big{|}\rv{w}_{k + u}^{(n)}{\big{|}^2}\right\}\sum\limits_{m = 1}^{{n_{\rm{t}}}} {\sum\limits_{l = 1}^L {\mathbb{E}\left\{\big{|}\rv{h}_{k,l}^{(mn)}{\big{|}^2}\right\}} } \mathbb{E}\left\{\big{|}\rv{s}_{k - l}^{(m)}{\big{|}^2}\right\} \\
& \quad
 + \mathbb{E}\left\{\big{|}\rv{w}_k^{(n)}{\big{|}^2}\right\}\sum\limits_{m = 1}^{{n_{\rm{t}}}} {\sum\limits_{l = 1}^L {\mathbb{E}\left\{\big{|}\rv{h}_{k + u,l}^{(mn)}{\big{|}^2}\right\}} } \mathbb{E}\left\{\big{|}\rv{s}_{k + u - l}^{(m)}{\big{|}^2}\right\} + \mathbb{E}\left\{\big{|}\rv{w}_k^{(n)}{\big{|}^2}\right\}\mathbb{E}\left\{\big{|}\rv{w}_{k + u}^{(n)}{\big{|}^2}\right\}. \label{fff}
\end{align}
\hrulefill
\vspace*{4pt}
\end{figure*}
\section{Conclusion}\label{section:conclusion}
{In this paper, we derived the \ac{da}- and \ac{nda}-\ac{crlb}s and \ac{da}- and \ac{nda}-\ac{mle}s for \ac{mds} in \ac{mimo} frequency-selective fading channel. Moreover,
a low-complexity \ac{nda}-\ac{mbe} for MISO and \ac{mimo} systems was
proposed. The \ac{nda}-\ac{mbe} employs the statistical moment-based approach and
relies on the second- and fourth-order statistics of
the received signal, as well as the LS curve-fitting optimization technique.
Compared to the existing DA estimators, the proposed \ac{nda}-\ac{mbe} provides higher system capacity due to absence of pilot. Also,
the substantial reduced-complexity enables the  proposed \ac{mbe} to exhibit good performance in the \ac{nda} scenarios, where the observation window can be selected large enough.
The \ac{nda}-\ac{mbe} does not require {\it{a priori}} knowledge of other parameters, such as the number of transmit antennas;
furthermore, the proposed  \ac{nda}-\ac{mbe} is robust to the time-frequency asynchronization.
When compared to the \ac{nda}-\ac{cce}, the \ac{nda}-\ac{mbe} exhibits better performance, and when compared to the low-complexity \ac{da}-\ac{mle}, it exhibits similar performance for high \ac{mds}s.
On the other hand, the derived \ac{da}-\ac{mle}'s performance is very close to the derived \ac{da}-\ac{crlb}
in \ac{mimo} frequency-selective channel even when the observation window is relatively small. Similarly, there is no significant performance gap between the derived \ac{nda}-\ac{mle}
and the \ac{nda}-\ac{crlb}.}

\appendices
\section{}\label{apd:proofI}

To obtain an explicit closed-form expression for $\kappa_u^{(n)} \buildrel \Delta \over =  \mathbb{E}\big{\{}|\rv{r}_k^{(n)}|^2|\rv{r}_{k+u}^{(n)}|^2\big{\}}$,
 we first write $\big{|}\rv{r}_k^{(n)}{\big{|}^2} = \rv{r}_k^{(n)}{\big{(}\rv{r}_k^{(n)}\big{)}^*}$ by employing \eqref{eq:TT1}, as in \eqref{uuu} at the top of next page.
 Then, $\big{|}\rv{r}_{k+u}^{(n)}{\big{|}^2}$ is straightforwardly calculated by replacing $k$ with $k+u$ in \eqref{uuu}, and
 $\big{|}\rv{r}_k^{(n)}\big{|}^2\big{|}\rv{r}_{k+u}^{(n)}\big{|}^2$ can be easily expressed in a summation form, which is omitted due to space constraints.

As fading is independent of the signal and noise, the statistical expectation in $\kappa_u^{(n)}$ can be decomposed into statistical expectations over the signal, fading, and noise distributions, respectively.
With independent and identically distributed transmitted symbols, $\rv{s}_k^{(m)}$, $k=1,\hdots,N$, and $u \ge L$, the symbols from the $m$th antenna contributed in $\rv{r}_k^{(n)}$, i.e., $\{\rv{s}_{k - l}^{(m)}\} _{l = 1}^L$, are different from those contributed in $\rv{r}_{k+u}^{(n)}$, i.e., $\{ \rv{s}_{k + u-l}^{(m)}\} _{l = 1}^L$; furthermore, by using that $\mathbb{E}\big{\{}{(\rv{s}_k^{({m})})^2}\big{\}} = 0$, $\mathbb{E}\big{\{}\rv{s}_k^{(m)}\big{\}} = 0$, and the linearity property of the statistical expectation, one obtains $\kappa_u^{(n)}$ as in \eqref{fff}.
Finally, with $\mathbb{E}\big{\{}|\rv{s}_k^{(m)}{|^2}\big{\}} = \sigma_{{\rm{s}}_{{m}}}^2$ and $\mathbb{E}\big{\{}|{\rv{w}}_k^{(n)}{|^2}\big{\}} = \sigma _{{\rm{w}}_n}^2$, \eqref{T44} is obtained. $\square$

{\section{}\label{ap2xc}
By employing \eqref{eq:TT9}, one can write
\begin{align}\label{eq:aw1} \nonumber
\Big{(}\mu_2^{(n)}\Big{)}^2 &\hspace{-0.2em}=\hspace{-0.2em}\Bigg{(}\sum\limits_{m = 1}^{{n_{\rm{t}}}} {\sum\limits_{l = 1}^L {\sigma _{{\rm{h}}_{(mn),l}}^2} } \sigma _{{\rm{s}}_{{m}}}^2 + \sigma _{{\rm{w}}_n}^2\Bigg{)}^2\hspace{-0.25em}=\hspace{-0.25em}\sum_{m=1}^{n_{\rm{t}}}\sum_{l=1}^{L}{\sigma_{{\rm{h}}_{(mn),l}}^4}{\sigma_{{\rm{s}}_{m}}^4} \\ \nonumber
&\quad +\sum\limits_{{m_1} = 1}^{{n_{\rm{t}}}} {\sum\limits_{{m_2} \ne {m_1}}^{n_{\rm{t}}} {\sum\limits_{{l} = 1}^L {\sigma _{{\rm{h}}_{(m_1n),l}}^2}{\sigma _{{\rm{h}}_{(m_2n),l}}^2} } } \sigma _{{\rm{s}}_{m_1}}^2\sigma _{{\rm{s}}_{m_2}}^2 \\ \nonumber
& \quad + \sum\limits_{m = 1}^{{n_{\rm{t}}}} {\sum\limits_{{l_1} = 1}^L {\sum\limits_{{l_2} \ne {l_1}}^L {\sigma _{{\rm{h}}_{(mn),l_1}}^2} \sigma _{{\rm{h}}_{(mn),l_2}}^2} } \sigma _{{{\rm{s}}_m}}^4 \\ \nonumber
& \quad + \sum\limits_{{m_1} = 1}^{{n_{\rm{t}}}} {\sum\limits_{{m_2} \ne {m_1}}^{{n_{\rm{t}}}} {\sum\limits_{{l_1} = 1}^L {\sum\limits_{{l_2} \ne {l_1}}^L {\sigma _{{\rm{h}}_{(m_1n),l_1}}^2\sigma _{{\rm{h}}_{(m_2n),l_2}}^2\sigma _{{\rm{s}}_{m_1}}^2\sigma _{{\rm{s}}_{m_2}}^2} } } } \\
& \quad + 2\sigma _{{\rm{w}}_n}^2\sum\limits_{m = 1}^{{n_{\rm{t}}}} {\sum\limits_{l = 1}^L {\sigma _{{\rm{h}}_{(mn),l}}^2} } \sigma _{{\rm{s}}_{{m}}}^2 +\sigma _{{\rm{w}}_n}^4.
\end{align}
Then, by subtracting $\big{(}\mu_2^{(n)}\big{)}^2$ in \eqref{eq:aw1} from $\kappa_u^{(n)}$ in \eqref{eq:TT8}, one obtains
\begin{equation}\label{eq:TT11o}
{\kappa_u^{(n)} -\Big{(}\mu_2^{(n)}\Big{)}^2}=\frac{J_0^2(2\pi {f_{\rm{D}}}{T_{\rm{s}}}u)}{\eta^{(n)} },
\end{equation}
where
$\eta^{(n)} ={1 \mathord{\left/
 {\vphantom {1 {}}} \right.
 \kern-\nulldelimiterspace} {\sum\limits_{m = 1}^{{n_{\rm{t}}}} {\sum\limits_{l = 1}^L {\sigma _{{\rm{h}}_{(mn),l}}^4} } \sigma _{{\rm{s}}_{{m}}}^4}}$.}
\section{}\label{ap3}
With independent fading, noise, and signal processes, by using \eqref{uuu} and then \eqref{set} and that $\mathbb{E}\{{(\rv{s}_k^{({m})})^2}\}=\mathbb{E}\{{\rv{s}_k^{(m)}}\} = \mathbb{E}\big{\{}\big{(}\rv{h}_{k,l}^{(mn)}\big{)}^2\big{\}} = \mathbb{E}\big{\{}{\rv{w}_k^{(n)}}\big{\}} = 0$, similar to Appendix  \ref{apd:proofI}, one obtains
\begin{align}\label{eq:C}
\mu_4^{(n)}&=\mathbb{E}\Big{\{}|\rv{r}_k^{(n)}|^4\Big{\}} =
2\sum\limits_{m = 1}^{{n_{\rm{t}}}} {\sum\limits_{l = 1}^L {\sigma _{{\rm{h}}_{(mn),l}}^4} } \sigma _{{{\rm{s}}}_{{m}}}^4
 \\ \nonumber
&\quad+2\sum\limits_{m = 1}^{{n_{\rm{t}}}} {\sum\limits_{l = 1}^L {\sigma _{{\rm{h}}_{(mn),l}}^4} } \left( {{\Omega _{\rm{s}}} - 1} \right)\sigma _{{\rm{s}}_{{m}}}^4 \\ \nonumber
&\quad+2\sum\limits_{{m_1} = 1}^{{n_{\rm{t}}}} {\sum\limits_{{m_2} \ne {m_1}}^{{n_{\rm{t}}}} {\sum\limits_{l = 1}^L \sigma _{{\rm{h}}_{(m_1n),l}}^2\sigma _{{\rm{h}}_{(m_2n),l}}^2\sigma _{{\rm{s}}_{m_1}}^2\sigma _{{\rm{s}}_{m_2}}^2  } } \\ \nonumber
&\quad
+ 2\sum\limits_{m = 1}^{{n_{\rm{t}}}} {\sum\limits_{{l_1} = 1}^L {\sum\limits_{{l_2} \ne {l_1}}^L {\sigma _{{\rm{h}}_{(mn),l_1}}^2\sigma _{{\rm{h}}_{(mn),l_2}}^2} } } \sigma _{{\rm{s}}_{{m}}}^4 \\ \nonumber
&\quad
 + 2\sum\limits_{{m_1} = 1}^{{n_{\rm{t}}}} {\sum\limits_{{m_2} \ne {m_1}}^{{n_{\rm{t}}}} {\sum\limits_{{l_1} = 1}^L {\sum\limits_{{l_2} \ne {l_1}}^L {\sigma_{{\rm{h}}_{(m_1n),l_1}}^2\sigma _{{\rm{h}}_{(m_2n),l_2}}^2\sigma _{s_{m_1}}^2\sigma _{{\rm{s}}_{m_2}}^2} } } } \\ \nonumber
&\quad  + 4\sigma_{{\rm{w}}_n}^2\sum\limits_{m = 1}^{{n_{\rm{t}}}} {\sum\limits_{l = 1}^L {\sigma _{{\rm{h}}_{(mn),l}}^2} } \sigma_{{\rm{s}}_{{m}}}^2 + 2\sigma_{{\rm{w}}_n}^4,
\end{align}
where $\Omega_{\rm{s}}$ is the fourth-order two-conjugate statistic for unit variance signal, which represents the effect of the modulation format.
Finally, by employing \eqref{eq:TT9} and \eqref{eq:C}, \eqref{eq:TT12x} is easily obtained. $\square$

%---------------------------------------------------------------------------%
%                            acknowledgment                                 %
%---------------------------------------------------------------------------%
%\section*{Acknowledgment}
%The authors are grateful to the anonymous reviewers and the
%Editor, Dr. Ruisi He, for their constructive comments.
\bibliographystyle{IEEEtran}
\bibliography{IEEEabrv,ref}

\end{document}